\documentclass[%
reprint,
superscriptaddress,
showpacs,
nofootinbib,
amsmath,amssymb,
aps,
prc,
showkeys,
twocolumn,
]{revtex4-1}

\usepackage{graphicx}
\usepackage{dcolumn}
\usepackage{bm}
\usepackage{hyperref}
\hypersetup{colorlinks=true, citecolor=blue, urlcolor=blue, linkcolor=blue}
\usepackage{color}
\usepackage{epsfig}
\usepackage{epstopdf}
\def\nuc#1#2{\relax\ifmmode{}^{#1}{\protect\text{#2}}\else${}^{#1}$#2\fi}

\usepackage{cancel}

\definecolor{Fred}{rgb}{1, 0, 1}

\begin{document}
  \newcommand {\nc} {\newcommand}
  \nc {\beq} {\begin{eqnarray}}
  \nc {\eeq} {\nonumber \end{eqnarray}}
  \nc {\eeqn}[1] {\label {#1} \end{eqnarray}}
  \nc {\etal} {\emph{et al.}}
  \nc {\eq} [1] {(\ref{#1})}
  \nc {\Eq} [1] {Eq.~(\ref{#1})}
  \nc {\Sec} [1] {Sec.~\ref{#1}}
  \nc {\Fig} [1] {Fig.~\ref{#1}}
  \nc {\IR} [1]{\textcolor{red}{#1}}
  \nc {\IB} [1]{\textcolor{blue}{#1}}

\title{Theoretical analysis of the extraction of neutron skin thickness from coherent $\pi^0$ photoproduction off nuclei}

\author{F.~Colomer}
 \affiliation{Physique Nucl\' eaire et Physique Quantique, Universit\'e Libre de Bruxelles (ULB), B-1050 Brussels}
 \affiliation{Institut f\"ur Kernphysik, Johannes Gutenberg-Universit\"at Mainz, 55099 Mainz, Germany}
  
\author{P.~Capel}
 \email{pcapel@uni-mainz.de}
  \affiliation{Institut f\"ur Kernphysik, Johannes Gutenberg-Universit\"at Mainz, 55099 Mainz, Germany}
 \affiliation{Physique Nucl\' eaire et Physique Quantique, Universit\'e Libre de Bruxelles (ULB), B-1050 Brussels}

\author{M.~Ferretti}
 \affiliation{Institut f\"ur Kernphysik, Johannes Gutenberg-Universit\"at Mainz, 55099 Mainz, Germany}

\author{J.~Piekarewicz}
\email{jpiekarewicz@fsu.edu}
\affiliation{Department of Physics, Florida State University, Tallahassee, FL 32306, USA}

\author{C.~Sfienti}
\email{sfienti@uni-mainz.de}
\affiliation{Institut f\"ur Kernphysik, Johannes Gutenberg-Universit\"at Mainz, 55099 Mainz, Germany}

\author{M.~Thiel}
 \email{thielm@uni-mainz.de}
 \affiliation{Institut f\"ur Kernphysik, Johannes Gutenberg-Universit\"at Mainz, 55099 Mainz, Germany}

\author{V.~Tsaran}
 \affiliation{Institut f\"ur Kernphysik, Johannes Gutenberg-Universit\"at Mainz, 55099 Mainz, Germany}

\author{M.~Vanderhaeghen}
\email{vandma00@uni-mainz.de}
\affiliation{Institut f\"ur Kernphysik, Johannes Gutenberg-Universit\"at Mainz, 55099 Mainz, Germany}

\date{\today}

\begin{abstract}

\begin{description}
\item[Background] Coherent $\pi^0$ photoproduction on heavy nuclei has been suggested as a reliable tool to infer neutron skin thicknesses.
To this aim, various experiments have been performed, especially on $^{208}$Pb.
\item[Purpose] We analyze the sensitivity of that reaction process to the nucleonic density, and especially to the neutron skin thickness, for $^{12}$C, $^{40}$Ca and $^{208}$Pb, for which reliable data exist, and on $^{116,124}$Sn, for which measurements have been performed in Mainz.
We study also the role played by the $\pi^0$-nucleus final-state interaction.
\item[Method] A model of the reaction is developed at the impulse approximation considering either plane waves or distorted waves to describe the $\pi^0$-nucleus scattering in the outgoing channel.
\item[Results] Our calculations are in good agreement with existing data, especially for $^{208}$Pb.
The sensitivity of the theoretical cross sections to the choice of the nucleonic density is small, and below the experimental resolution.
\item[Conclusions] Coherent $\pi^0$ photoproduction is mostly an isoscalar observable that bares no practical sensitivity to the neutron skin thickness.
To infer that structure observable it should be coupled to other reaction measurements, such as electron scattering, or by comparing experiments performed on isotopes of the same chemical element.
\end{description}
\end{abstract}

\maketitle

\section{\label{sec:intro}Introduction}

One of the major challenges of the 21st century in nuclear physics is the determination of the nuclear equation of state (EoS) and more 
specifically the density dependence of the symmetry energy\,\cite{Tsa12,Hor14}. The equation of state governs the structure of the 
densest objects in the universe, from the bulk properties of heavy nuclei up to the structure of neutron stars. This exciting field of 
research has led to the development of a large range of both observational and experimental techniques to constrain the EoS from 
both terrestrial measurements and multi-messenger astronomy\,\cite{Tsa12,Hor14,Thi19}.

The asymmetry term of the EoS quantifies the energy cost of converting symmetric nuclear matter, where the number of neutrons equals 
that of protons, into pure neutron matter. The structure of neutron stars is strongly sensitive to the asymmetry term---and especially to its 
density dependence---because it provides the dominant contribution to the baryonic pressure in the vicinity of nuclear matter saturation 
density.  Indeed, it has been argued that the pressure near twice saturation density sets the overall scale for stellar radii\,\cite{Lattimer:2006xb}.
In heavy nuclei with a significant neutron excess, the development of a \emph{neutron skin} emerges from a competition between surface
tension and the density dependence of the symmetry energy. In the context of the liquid-drop model, surface tension favors the creation
of an incompressible drop with the smallest possible area. In contrast, the symmetry energy favors moving the excess neutrons from the 
core---where the asymmetry term is large---to the surface, where the asymmetry term is small. The difference between the value of the
asymmetry term at the center of the nucleus relative to its value at the surface is denoted by $L$. As such, the \emph{asymmetry pressure} 
$L$ controls both the thickness of the neutron skin and the radius of a neutron star\,\cite{Horowitz:2000xj,Horowitz:2001ya}.

In the laboratory, the most direct way to constrain the asymmetry pressure is to measure the thickness of the neutron skin in heavy 
nuclei\,\cite{Tsa12,Hor14,Thi19}. The neutron skin thickness is defined as the difference between the root-mean-square radius of the
neutron density relative to that of the proton. That is,
\begin{equation}
 R_{\rm skin} = R_{n}-R_{p}=\sqrt{\langle r^2_n\rangle} - \sqrt{\langle r^2_p\rangle}. 
 \label{e2}
\end{equation}
For this study, the choice of $^{208}$Pb is optimal. It is stable, which makes it readily available for laboratory experiments, it exhibits a large 
neutron-proton asymmetry, and its heavy, doubly-magic nature makes it amenable to mean-field calculations. 

Various experimental techniques have been implemented to extract the neutron skin thickness of $^{208}$Pb\,\cite{Tsa12,Hor14,Thi19}. 
Parity violating elastic electron scattering is widely regarded as the cleanest experimental technique from which $R_{\rm skin}^{208}$ can 
be inferred. The original lead radius experiment (PREX) infers $R_{\rm skin}^{208}\!=\!0.33^{+0.16}_{-0.18}$\,fm,\cite{PREXI}, a value that 
has since been refined by the recent PREX-2 campaign to $R_{\rm skin}^{208}\!=\!0.283\pm0.071$\,fm\,\cite{PREXII}. Elastic proton 
scattering experiments suggest a thinner neutron skin with a significantly smaller uncertainty, namely, 
$R_{\rm skin}^{208}\!=\!0.211^{+0.054}_{-0.063}$\,fm\,\cite{Zen10}. Antiprotonic atoms measurements have inferred an even smaller 
neutron skin of $R_{\rm skin}^{208}\!=\!0.16\pm0.02$(stat.)$\pm0.04$(syst.)\,fm\,\cite{Trz01,Klo07}. To date, the smallest estimate of 
the neutron skin thickness of $^{208}$Pb, $R_{\rm skin}^{208}\!=\!0.156^{+0.025}_{-0.021}$\,fm, was obtained from a measurement of 
the electric dipole polarizability\,\cite{Tam11}, a quantity that has been shown to be strongly correlated to the neutron skin 
thickness\,\cite{Reinhard:2010wz}. Although not fully inconsistent, these estimates spread over a broad range of values, leading to 
significant uncertainties in the inferred contribution of the asymmetry term to the nuclear EoS. The large spread is generated from the 
significant model dependence involved in the extraction of $R_{\rm skin}^{208}$ from the various experiments\,\cite{Thi19}. This
is particularly true in the case of hadronic reactions that suffer from large systematic errors. To consistently compare this suite of 
laboratory experiments with recent astronomical observations it is imperative to properly quantify the systematic and model 
uncertainties. In this context, we note that the PREX error is dominated by \emph{statistical} errors. 

In this work we focus on another experimental method that has been suggested to pin down the value of the neutron skin thickness: 
the coherent $\pi^0$ photoproduction. In such a reaction, an incident photon generates through its interaction with the target nucleus a 
neutral pion that can then be easily detected by its dominant $2\gamma$ decay channel\,\cite{PhysRev.127.1772, RevModPhys.30.456}.
At the level of the plane-wave impulse approximation (PWIA), the cross section is directly proportional to the Fourier transform of the nuclear 
density\,\cite{NPA660}. An accurate measurement would give access to the entire baryon density and hence to an estimate of the 
neutron skin thickness, given that the proton density is well known from decades of electron-scattering experiments\,\cite{Fri95}.
Measurements of coherent $\pi^0$ photoproduction on different targets and at various photon energies already 
exist\,\cite{KRUSCHE2002287,PhysRevLett112_242502}. In one such analysis, Tarbert \etal\ have inferred a rather thin neutron skin 
thickness for $^{208}$Pb with a very small uncertainty: 
$R_{\rm skin}^{208}\!=\!0.15\pm0.03(\mathrm{stat.})^{+0.01}_{-0.03}(\mathrm{sys.})$\,fm \cite{PhysRevLett112_242502}. Given the
hadronic character of the production channel, the small systematic error seems unrealistic. Indeed,  a theoretical analysis suggests 
that the uncertainty claimed by Tarbert \etal\ is significantly underestimated because of second-order effects that were not included 
in the model used to analyze the data\,\cite{Mil19}.

Here we follow a complementary approach by studying in detail the sensitivity (or lack-thereof) of the $\pi^0$-photoproduction 
cross section to the choice of nuclear density. Moreover, we also analyze the role played by the interaction between the pion and 
the nucleus in the final state. To reach this goal, we have written a reaction code based on the impulse approximation (IA)\,\cite{NPA660,PLM98}, 
which is used to analyze the experimental data\,\cite{KRUSCHE2002287,PhysRevLett112_242502}. In the impulse approximation, one
assumes that the elementary $\gamma N \to \pi^0 N$ interaction remains unchanged in the nuclear medium. The approach proposed
here can examine the sensitivity of the cross section to various choices of neutron densities as well as to the impact of $\pi^0$-nucleus 
final-state interactions by performing both plane-wave  (PWIA) and distorted wave (DWIA) calculations.

The model is briefly described in \Sec{sec:theory} and its predictions are confronted against existing experimental data on $^{12}$C, $^{40}$Ca 
and $^{208}$Pb\,\cite{KRUSCHE2002287} in \Sec{sec:CCaPb}. In that section, the sensitivity of the theoretical cross sections to the choice 
of nuclear density is carefully analyzed and the role played by the $\pi^0$-nucleus interaction in the final state is investigated. The same model
is then applied to two tin isotopes---$^{116}$Sn and $^{124}$Sn---for which the $\pi^0$-photoproduction cross section has been measured at 
Mainz. Section~\ref{sec:Sn} summarizes our predictions for those two isotopes and estimates the experimental precision that must be reached to 
enable us to infer information about the evolution of the neutron skin thickness along the tin isotopic line.  A brief summary and conclusions 
are presented in \Sec{sec:conclusion}.

\section{\label{sec:theory} Coherent $\pi^0$-photoproduction at the impulse approximation}

We consider the reaction induced by a photon $\gamma$ on a nucleus $A$ that produces a neutral pion $\pi^0$ while leaving 
the nucleus in its initial ground state. This coherent $\pi^0$-photoproduction process reads
\begin{equation}
\gamma + A \to \pi^0 + A. \label{e1}
\end{equation}
To model this reaction, we consider only the dominant one-body mechanism of photoproduction in which the neutral pion is 
produced on a single nucleon of the target $A$. Two-body processes, such as the one in which a charged pion is produced 
on a first nucleon and then charge exchanges into a neutral pion onto another nucleon, are neglected at the level of the 
impulse approximation adopted here\,\cite{NPA660,PLM98}.

\begin{figure}[htb]
	\includegraphics[width=\columnwidth]{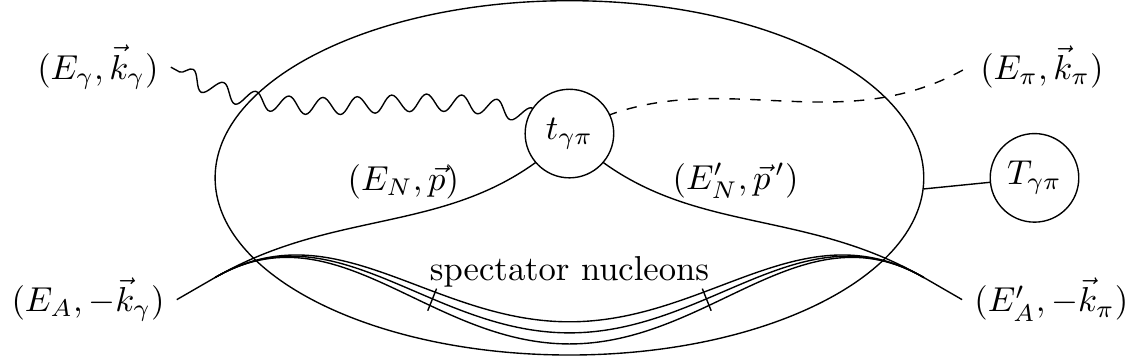}
	\caption{\label{f1} Kinematics of the pion-photoproduction reaction on a nucleus in the $\gamma$-$A$ ($\equiv\pi$-$A$) rest frame described at the impulse approximation. The four-vectors of the photon, the nucleus before the reaction, the pion and the nucleus after the reaction correspond to the indices $\gamma$, $A$, $\pi$ and $A'$, respectively. Inside the nucleus, the four-vectors of the struck nucleon before and after the reaction read $(E_N,\vec{p})$ and $(E'_N,\vec{p}^{\,\prime})$.}
\end{figure}

A description of this reaction at the IA, including its relevant kinematics, is provided in \Fig{f1}: The initial photon of energy 
$E_\gamma$ and momentum $\vec{k}_\gamma$ within the $\gamma$-$A$ center-of-momentum frame collides with a 
nucleon $N$ of energy $E_N$ and momentum $\vec{p}$, producing the outgoing neutral pion $\pi$ of energy $E_\pi$ and 
momentum $\vec{k}_\pi$. The remaining $A\!-\!1$ nucleons in the target are treated as spectators. The process is coherent
because the total amplitude is obtained as the sum over the individual nucleon amplitudes. Further, at the level 
of the impulse approximation, it is assumed that the elementary on-shell amplitude is unchanged in the nuclear medium; 
that is, as far as the elementary amplitude is concerned, the nucleons are treated as free. In the particular case of a 
spin zero nucleus, the photoproduction amplitude for a photon of polarization $\lambda$ reads\,\cite{NPA660}
\beq
U^{(\lambda)}_{\gamma \pi}(\vec{k}_\pi,\vec{k}_\gamma) &=& \frac{-i\lambda e^{i\phi\lambda}}{\sqrt{2}} \mathcal{W}_A \left[f^p_2(\vec{k}_\pi,\vec{k}_\gamma)  \mathcal{F}[\rho_p](q) \right.\nonumber\\
& &+ \left.f^n_2(\vec{k}_\pi,\vec{k}_\gamma) \mathcal{F}[\rho_n](q)\right],
\eeqn{UPWIA}
where $\vec{q}=\vec{k}_\pi-\vec{k}_\gamma$ is the momentum transferred during the process, $\phi$ is the azimuthal angle of the outgoing pion relative to the incoming photon, $\mathcal{W}_A$ is a phase-space factor that we choose identical to the one defined in Ref.~\cite{NPA660}, $f_2^N$ are the elementary photoproduction amplitudes on a single nucleon, either proton ($N=p$) or neutron ($N=n$), and $\mathcal{F}[\rho_N]$ are the Fourier transforms of the nucleonic densities
\begin{equation}
\mathcal{F}[\rho_N](q) = \int e^{i\vec{q}\cdot\vec{r}}\rho_N(\vec{r}) d\vec{r}. \label{FTdens}
\end{equation}

The elementary photoproduction amplitudes $f_2^N$ are calculated in the $\gamma$-$A$ ($\equiv\pi$-$A$) rest frame as in Ref.~\cite{NPA660}
\begin{align}
f_2^N(\vec{k}_\pi,\vec{k}_\gamma) = \frac{k_\pi k_\gamma}{\tilde{k}_\pi \tilde{k}_\gamma} \left(1+\frac{A-1}{2A}(\alpha_{\gamma}+\alpha_{\pi})\right) & \notag\\
\times \quad \mathcal{F}_2^N(\tilde{\vec{k}}_\pi,\tilde{\vec{k}}_\gamma ; W_{\gamma N}), 
\end{align}
where the tilde on momenta denote momenta calculated in the $\gamma$-$N$ ($\pi$-$N$) rest frame as opposed to the $\gamma$-$A$ ($\pi$-$A$) frame and $\theta$ is the scattering angle in the $\pi$-$A$ rest frame.
The coefficients $\alpha_{\gamma(\pi)}$ are defined by the Lorentz transformations between the $\gamma$-$A$ ($\pi$-$A$) and $\gamma$-$N$ ($\pi$-$N$) frames as
\begin{equation}\label{LorTrans}
\tilde{\vec{k}}_{\gamma(\pi)} = \vec{k}_{\gamma(\pi)} + \alpha_{\gamma(\pi)} \vec{P}
\end{equation}
where $\vec{P}=\vec{k}_\gamma + \vec{p}\; (= \vec{k}_\pi + \vec{p}^{\,\prime})$ is the total momentum of the $\gamma$-$N$ ($\pi$-$N$) system evaluated in the $\gamma$-$A$ ($\pi$-$A$) rest frame. Accordingly, $\alpha_{\gamma(\pi)}$ is given by
\begin{equation}\label{alpha}
\alpha_{\gamma(\pi)} = \frac{1}{W_{\gamma N (\pi N)}} \left(\frac{\vec{P}\cdot\vec{k}_{\gamma(\pi)}}{E_{\gamma(\pi)} + E^{(\prime)}_{N} + W_{\gamma N(\pi N)}} - E_{\gamma(\pi)}\right), 
\end{equation}
with  $W_{\gamma N(\pi N)}$ the $\gamma$-$N$ (respectively $\pi$-$N$) relative energy.
In this work, the standard CGLN amplitudes $\mathcal{F}_2^N$ (Chew, Goldberger, Low, and Nambu \cite{CGLN57}) are taken from the database MAID  \cite{MAID}.
%
%

The coherent $\pi^0$-photoproduction cross section reads
\begin{equation}
\frac{d\sigma^{\gamma\pi}}{d\Omega} = \frac{k_\pi}{k_\gamma}\frac{1}{2}\sum_\lambda\left|T_{\gamma\pi}^{(\lambda)}\right|^2 \label{xsec}
\end{equation}
where $T_{\gamma\pi}^{(\lambda)}$ is the $T$ matrix for this process with an impinging photon of polarisation $\lambda$.
At the plane wave impulse approximation (PWIA), the final state interaction between the emitted pion and the nucleus is neglected.
The $T$-matrix then simply equals the $\pi^0$-photoproduction amplitude $U^{(\lambda)}_{\gamma \pi}$ \eq{UPWIA}
\begin{equation}
T_{\gamma\pi}^{(\lambda)} \stackrel{\text{\tiny PWIA}}{=} U^{(\lambda)}_{\gamma \pi}
\end{equation}

Because the photoproduction amplitude $U^{(\lambda)}_{\gamma \pi}$ depends linearly on the Fourier transforms $\mathcal{F}[\rho_N]$ of the nucleonic densities, at the PWIA, the cross section for $\pi^0$-photoproduction Eq.~\eqref{xsec} should gives us direct access to the nuclear density.
In combination with charge density obtained, e.g., through electron elastic scattering measurements \cite{Fri95}, this observable could thus be used to infer information about the neutron distribution, and hence about the neutron skin thickness.

Final-state interactions have been shown to play a significant role in this reaction \cite{NPA660,KRUSCHE2002287,PLM98}.
They can be taken into account at the distorted wave impulse approximaton (DWIA), within which the $T$-matrix reads
\begin{equation}
T^{(\lambda)}_{\gamma \pi} \stackrel{\text{\tiny DWIA}}{=} U^{(\lambda)}_{\gamma \pi} - 4\pi \frac{A-1}{A} T_{\pi A} G_0 U^{(\lambda)}_{\gamma \pi} \label{eqTgammapi}
\end{equation}
The first term corresponds to the PWIA term discussed above [see \Eq{UPWIA}], while the second accounts for the interaction of the $\pi^0$ with the nucleus in the exit channel.
The $T$ matrix $T_{\pi A}$ describes the $\pi$-$A$ scattering and is obtained by solving the Lippmann-Schwinger equation
\begin{equation}\label{eqLS}
T_{\pi A} = U_{\pi A} - 4\pi \frac{A-1}{A} U_{\pi A} G_0 T_{\pi A}
\end{equation}
where $U_{\pi A}$ is an optical potential that simulates the $\pi$-$A$ interaction.
We solve \Eq{eqLS} through a partial-wave expansion following Ref.~\cite{PhysRevC.31.1349}.

In both Eqs.~\eq{eqTgammapi} and \eq{eqLS}, $G_0$ is the free pion-nucleus propagator
\begin{equation}
G_0(k) = \frac{1}{2\mathcal{M}(k)}\frac{1}{E(k_0)-E(k)+i\epsilon}
\end{equation}
where $\mathcal{M}$ is a relativistic equivalent of the $\pi$-$A$ reduced mass and is chosen as in Ref.~\cite{NPA660}.

The shape of the optical potential $U_{\pi A}$ is chosen from Ref.~\cite{PhysRevC25_952} and will be later referred to as the MSU potential.
In momentum space, its matrix elements for a pion scattered from $\vec{k}_\pi$ to $\vec{k}'_\pi$ read
%
\beq
U_{\pi A}(\vec{k}'_\pi,\vec{k}_\pi) &= &p_1 \bar{b}_0 \mathcal{F}[\rho](q_\pi) + p_2 B_0\mathcal{F}[\rho^2](q_\pi) \nonumber\\
& - & \left\{p'_1 c_0 \mathcal{F}[\rho](q_\pi) + p'_2 C_0 \mathcal{F}[\rho^2](q_\pi) \right\} q_\pi^2\nonumber \\
& + & (\vec{k}_\pi\cdot\vec{k}'_\pi) \mathcal{L}(q_\pi),
\eeqn{eUpA}
where $\rho=\rho_p+\rho_n$ is the nucleonic density of $A$, $\vec{q}_\pi = \vec{k}'_\pi - \vec{k}_\pi$ is the transferred momentum and the function $\mathcal{L}$ is defined as 
\begin{equation}
\mathcal{L}(q) = \mathcal{F}\left[\frac{L}{1+\frac{4\pi}{3}\lambda_{\rm LLEE} L}\right](q) \label{L(q)},
\end{equation}
where $\lambda_{\rm LLEE}$ is the Lorenz-Lorentz-Ericson-Ericson parameter, first introduced in Ref.~\cite{ERICSON1966323}, and
\begin{equation} 
L(r) = p_1^{-1} c_0 \rho(r) + p_2^{-1} C_0 \rho^2(r)
\end{equation}
In all these expressions, $p^{(\prime)}_1$ and $p^{(\prime)}_2$ are kinematic factors whose expressions are 
\begin{align}
	p_1 = \frac{1+\varepsilon}{1+\varepsilon/A} \quad&\quad p_2 = \frac{1+\varepsilon/2}{1+\varepsilon/2A} \\
	p'_1= \frac{1}{2} (1-p^{-1}_1) \quad&\quad p'_2 = \frac{1}{2} (1-p^{-1}_2)
\end{align}
where $\varepsilon = \sqrt{k_{\pi,0}^2+m_\pi^2}/m_N$, with $k_{\pi,0}$ the on-shell pion momentum.
The values of the complex parameters $\bar{b}_0$, $c_0$, $B_0$, $C_0$ and $\lambda_{\rm LLEE}$ are chosen identical to the ones of the set E in Ref.~\cite{PhysRevC25_952}.
These values have been fitted to reproduce elastic-scattering cross sections of charged pions measured at a laboratory kinetic energy of 50~MeV on several targets from $^{12}$C to $^{208}$Pb.
This choice is therefore rather well suited for the purpose of this work.


%

\section{\label{sec:CCaPb}Theoretical analysis of $\pi^0$ photoproduction on $^{12}\rm C$, $^{40}\rm Ca$ and $^{208}\rm Pb$}

Krusche \textit{et al.} have measured the cross section for coherent $\pi^0$ photoproduction on several targets---including $^{12}$C, $^{40}$Ca, 
and $^{208}$Pb---at an incoming photon energy of 200~MeV \cite{KRUSCHE2002287}. This constitutes an excellent set of data to which to compare 
the predictions of the model presented in \Sec{sec:theory}, explore the sensitivity of the calculations to the choice of the target densities, and analyze 
the impact of final state interactions. For that last goal, we use the MSU potential\,\cite{PhysRevC25_952}, neglecting the small energy difference 
between its range of validity and the actual experimental conditions of Ref.\,\cite{KRUSCHE2002287}. Moreover, given that the neutron densities in
both $N\!=\!Z$ nuclei $^{12}$C and $^{40}$Ca are expected to follow closely the corresponding proton density, these two nuclei provide an excellent 
testing ground for assessing the role of final state interactions. 

\subsection{Nucleonic densities}\label{densities}

\begin{figure}[htb]
	\includegraphics[width=\columnwidth]{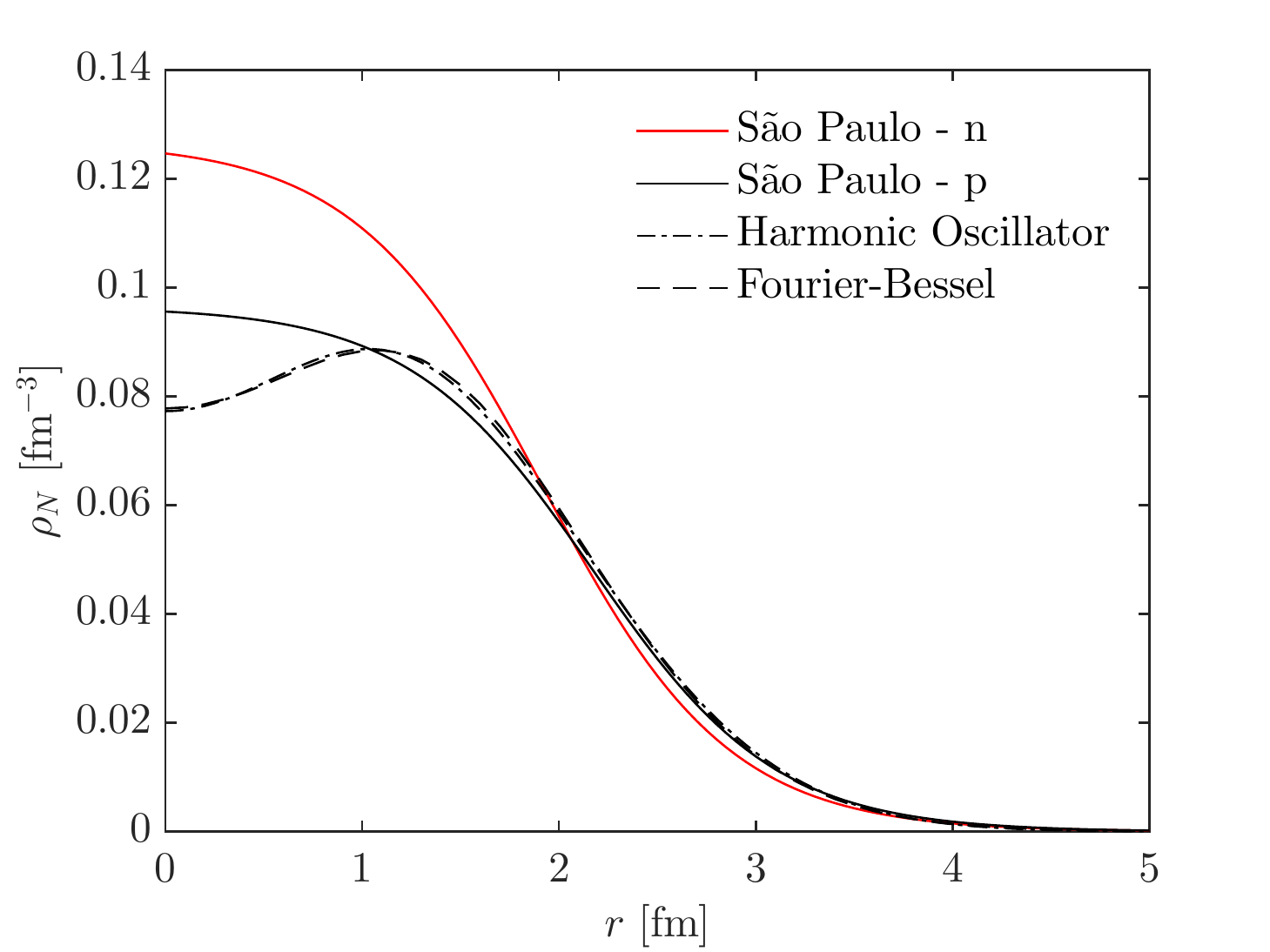}
	\caption{\label{f2C} Nucleonic densities of $^{12}$C considered in the calculations of the $\pi^0$ photoproduction: the S\~ao Paulo parametrization (solid lines) \cite{PhysRevC.66.014610}, densities derived from a harmonic-oscillator mean field (dash-dotted line) \cite{DJDV74}, and those inferred from electron elastic scattering using a Fourier-Bessel expansion (dashed line) \cite{DREHER1974219}. The S\~ao Paulo density differs for protons (black solid line) and neutrons (red solid line), whereas in the other cases, we assume $\rho_n=\rho_p$.}
\end{figure}

For our calculations on $^{12}$C, we consider three different choices of nuclear densities, which are displayed in Fig.~\ref{f2C} as a function of the 
radial distance $r$. The first one, shown in solid lines (with the proton density in black and the neutron density in red), is the phenomenological density 
of the S\~ao Paulo group \cite{PhysRevC.66.014610}. It exhibits a simple Fermi-Dirac shape whose parameters have been adjusted by the known
charge distributions inferred from electron-scattering experiments as well as theoretical densities derived from mean-field models across the entire 
nuclear chart. The second one is a density obtained by considering the nucleons to be bound in a mean-field harmonic oscillator (HO) potential well 
(dash-dotted line)\,\cite{DJDV74}. The third density, plotted with dashed lines, corresponds to the charge distribution fitted to reproduce 
electron-scattering data and parametrized with a Fourier-Bessel (FB) expansion\,\cite{DREHER1974219}. For the HO and FB densities, we assume
that $\rho_n\!=\!\rho_p$, which is sensible for this light, stable, $N\!=\!Z$ nucleus. Note that accurately-calibrated mean-field models predict a very 
small and negative neutron skin thickness in $^{12}$C because of the Coulomb repulsion among protons. Both HO and FB densities exhibit nearly 
identical radial dependences, in particular in the exponential tail. On the contrary, the S\~ao Paulo prediction differs significantly from the other two.
Its $p$ and $n$ densities are very different from one another, a fact that is difficult to understand from the perspective of mean-field models. Whereas
the proton density remains close to the HO and FB predictions beyond 1\,fm, the $n$ density decays too fast at large $r$, leading to an unrealistic 
\emph{negative} neutron skin thickness. Nevertheless, we consider the S\~ao Paulo model to assess the sensitivity of the coherent process to the 
choice of densities. Moreover, this parametrization is available throughout the entire nuclear chart, which enables us to compare the calculations 
performed on different targets.

\begin{figure}[htb]
	\includegraphics[width=\columnwidth]{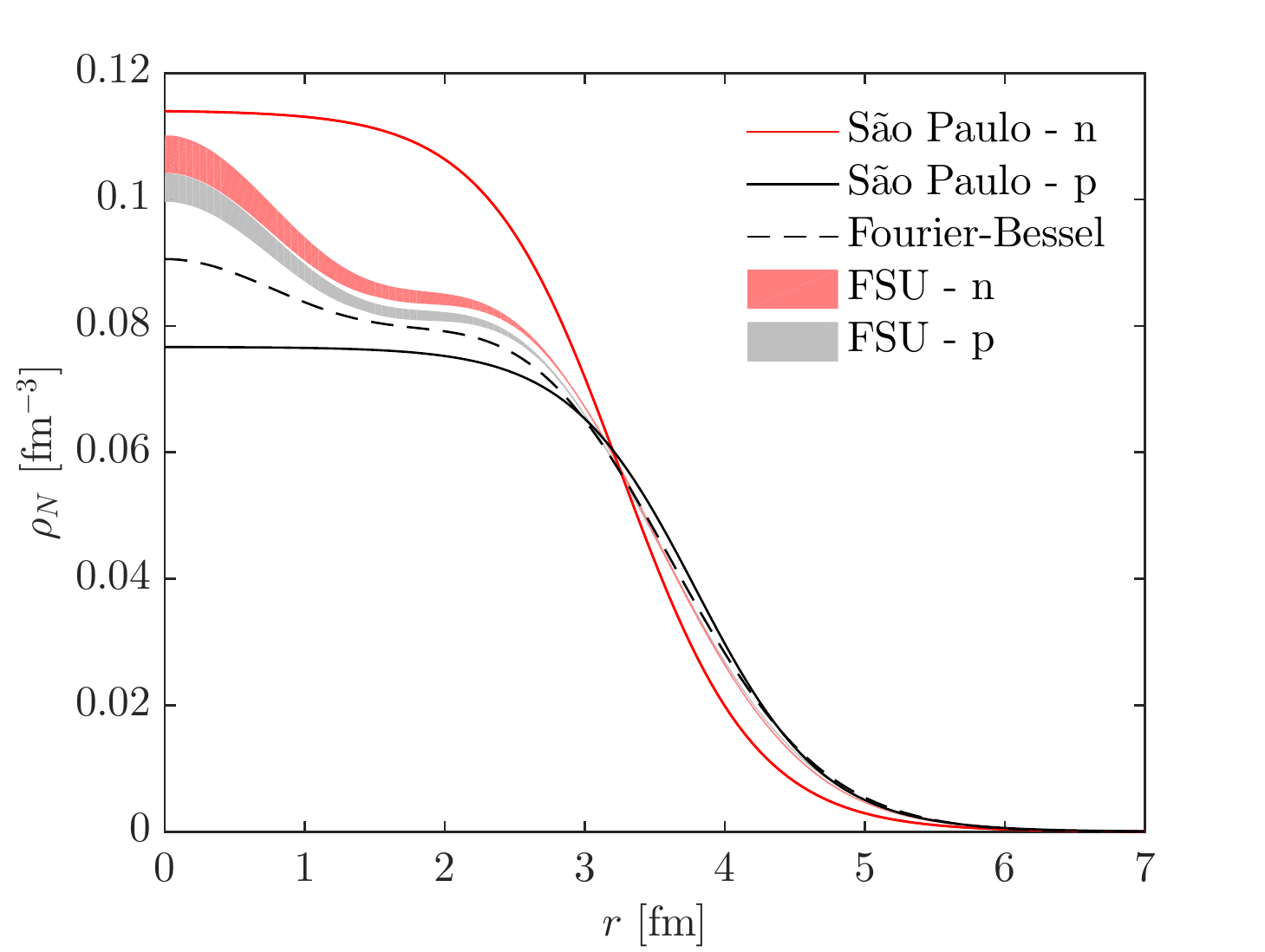}
	\caption{\label{f2Ca} 
	Nucleonic densities of $^{40}$Ca: S\~ao Paulo parametrization (solid lines) \cite{PhysRevC.66.014610}, Fourier-Bessel expansion (dashed line) \cite{DREHER1974219}, and FSU relativistic mean filed calculations (bands) \cite{PhysRevLett.95.122501}. In the case of the Fourier-Bessel case, we 
	assume $\rho_n\!=\!\rho_p$.}
\end{figure}

For the $^{40}$Ca target, in addition to the S\~ao Paulo and Fourier-Bessel parametrizations, we add predictions from relativistic mean field calculations 
performed within the Florida State University (FSU) model of Ref.\,\cite{PhysRevLett.95.122501}. All three choices are displayed in \Fig{f2Ca}. We follow 
the same line-type convention as in \Fig{f2C} for the first two. The FSU calculations have been performed with different choices of nonlinear coupling 
between the isoscalar and the isovector mesons to modify the neutron skin thickness of the nucleus without changing its isoscalar properties. Because 
these predictions are very close to one another, they are presented in \Fig{f2Ca} as bands: red for neutrons and grey for protons.

In this case, the differences between the three density choices is larger than for $^{12}$C. As such, it should help us assess the influence of this 
nuclear-structure observable---and in particular of the neutron skin thickness---on the $\pi^0$-photoproduction reaction. As for $^{12}$C, the 
S\~ao Paulo parametrization\,\cite{PhysRevC.66.014610} predicts a very compact neutron density that decays much faster at large distances 
than $\rho_p$ and than the other choices of densities. All FSU parametrizations predict very similar proton and neutron densities, with a slightly 
more compact $\rho_n$\,\cite{PhysRevLett.95.122501}, leading to a slightly negative neutron skin thickness. Given that the charge radius of 
$^{40}$Ca was incorporated into the calibration of the FSU model, its predictions at intermediate to large distances are close to the ones 
obtained from the Fourier-Bessel expansion\,\cite{DREHER1974219}. With the exception of the Fourier-Bessel expansion, for which we again
assume that $\rho_n\!=\!\rho_p$, all models predict negative neutron skin thicknesses. The first line of Table~\ref{tab1} lists these predictions. 
The negative value of the neutron skin thickness of $^{40}$Ca is not uncommon given that the Coulomb repulsion pushes protons out to
the surface: this has also been found by Hagen \textit{et al.} in their recent \emph{ab initio} coupled cluster calculation \cite{Hag16,Hag16p}.
However, the value obtained with the S\~ao Paulo parametrization is unrealistically large. Nevertheless, such a large spread in the model 
predictions will enable us to test the sensitivity of $\pi^0$ photoproduction to the neutron density.

\begin{table}[h!]
	\caption{\label{tab1} Neutron skin thicknesses for $^{40}$Ca and $^{208}$Pb predicted by the S\~ao Paulo parametrization \cite{PhysRevC.66.014610} and the FSU relativistic mean field calculations \cite{PhysRevLett.95.122501}.}
	\centering
	\begin{tabular}{c | c  c }\hline\hline
			Model & São Paulo & FSU \\
			\hline
			$^{40}$Ca & $-0.301$~fm & $[-0.051, -0.049]$~fm\\
			$^{208}$Pb & \ \ 0.101~fm & $[0.176, 0.286]$~fm\\ \hline\hline
	\end{tabular}
\end{table}

\begin{figure}[t]
	\includegraphics[width=\columnwidth]{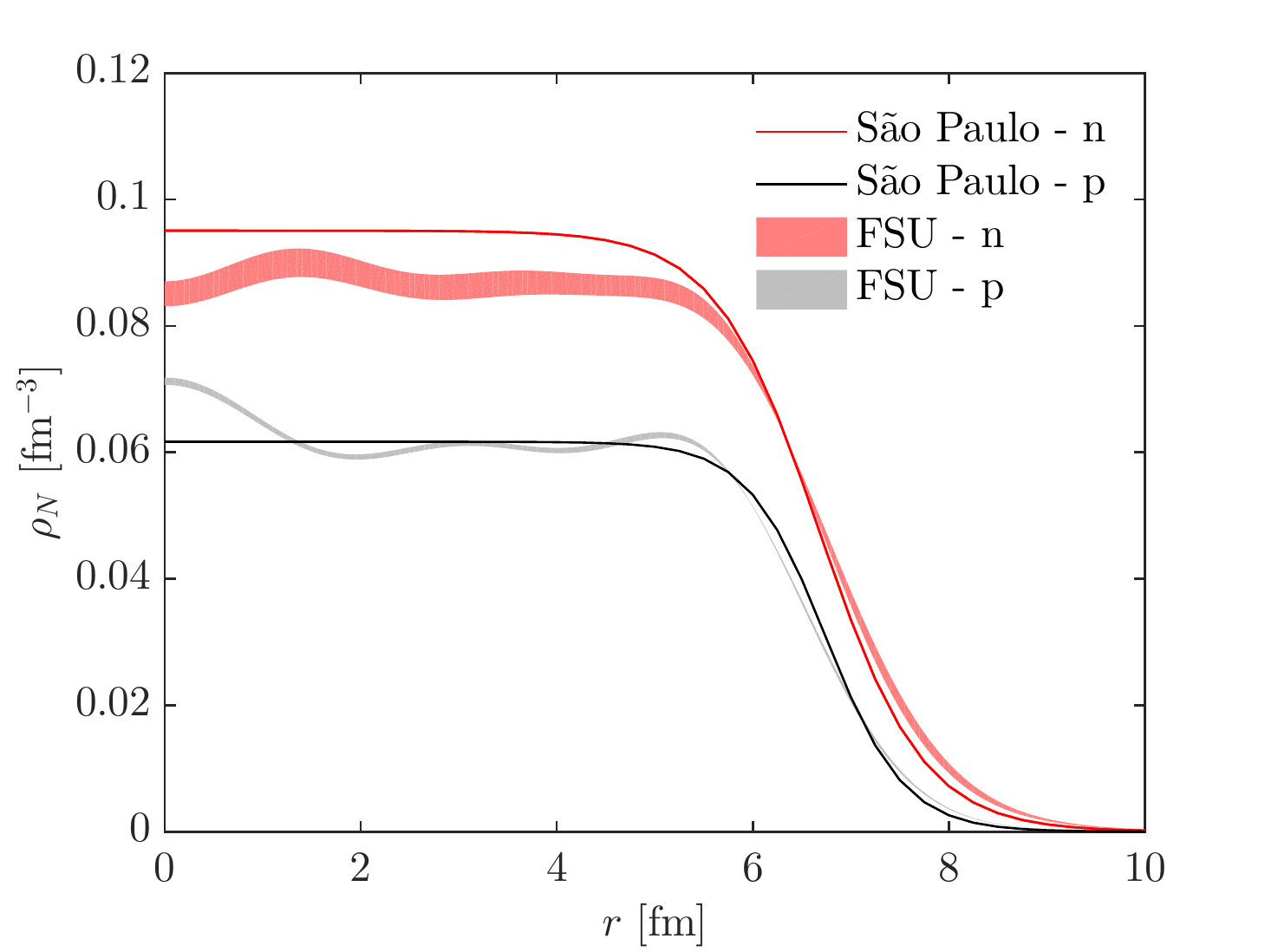}
	\caption{\label{f2Pb} 
	Nuclear densities of $^{208}$Pb: S\~ao Paulo parametrization (solid lines) \cite{PhysRevC.66.014610} and FSU relativistic mean field calculations (bands) \cite{PhysRevLett.95.122501}.
	The neutron densities are shown in red and the proton ones in black/grey.}
\end{figure}

For the heavy $^{208}$Pb target, the sole charge density inferred from electron scattering is of little help in estimating the neutron density. Accordingly, 
only the S\~ao Paulo and FSU densities are considered; they are displayed in Fig.~\ref{f2Pb} as solid lines and bands, respectively. As for $^{40}$Ca, 
those bands correspond to the most extreme variation in the FSU model that still predicts the correct $^{208}$Pb binding energy---and in general all
observables dominated by the isoscalar sector. In this case, the agreement between both models is better than for $^{40}$Ca, although the S\~ao Paulo 
neutron density continues to decay faster relative to the FSU predictions, leading to a neutron skin thickness of about 0.1\,fm, that is inconsistent with
the extracted value from PREX-2. The significant spread of nearly 0.1\,fm in the FSU predictions for $^{208}$Pb shown in Table~\ref{tab1} will be 
particularly valuable for our analysis of the $\pi^0$-photoproduction reaction---and the unrealistically small neutron skin thickness predicted by the 
S\~ao Paulo parametrization will enable us to extend our tests to even more extreme nucleonic densities.
  
\subsection{PWIA and DWIA calculations of $\pi^0$ photoproduction}

\begin{figure*}
        \includegraphics[trim={0.55cm 0.05cm 1.39cm 0.78cm}, clip, height=0.85\columnwidth]{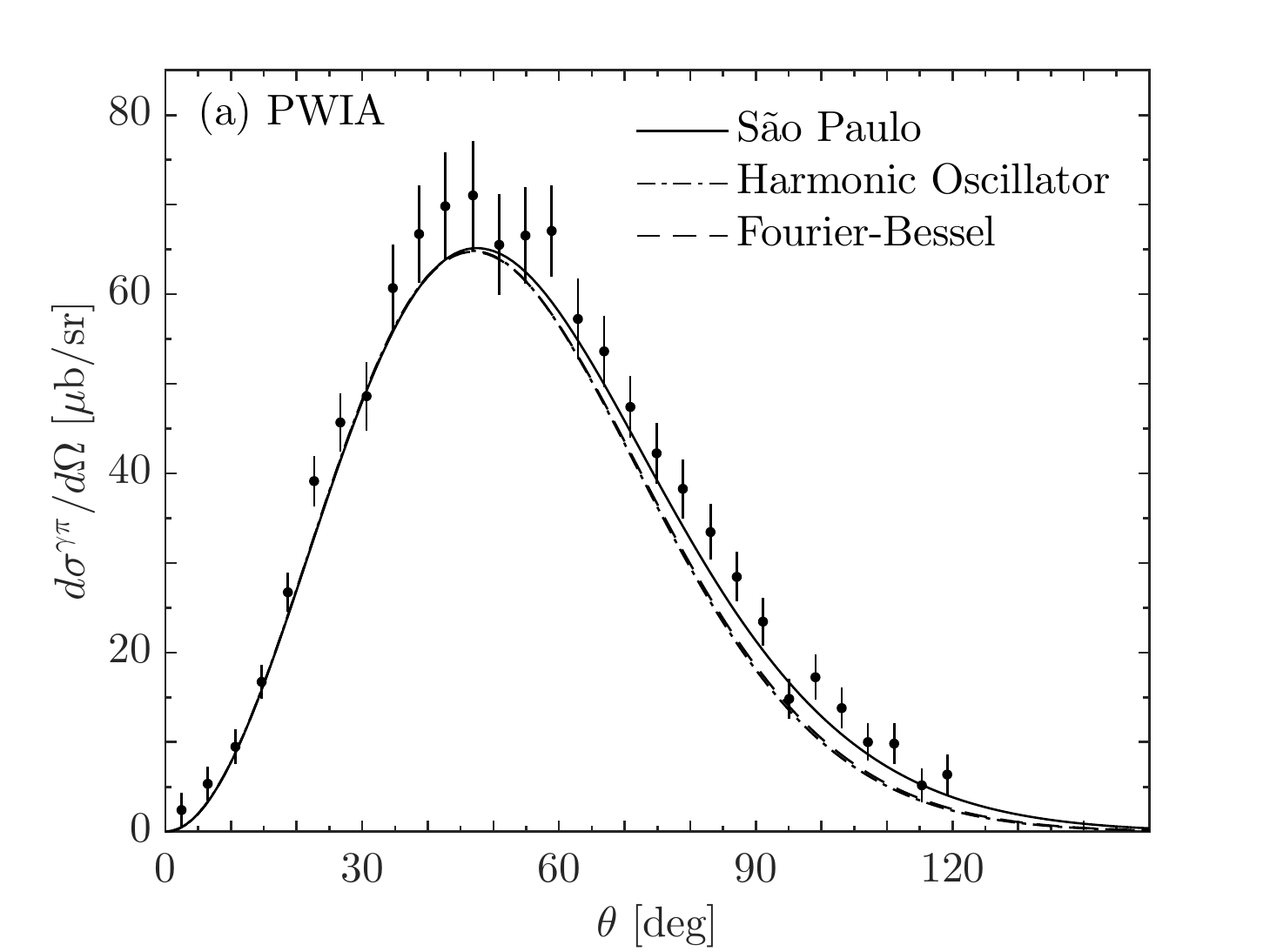}\includegraphics[trim={1.93cm 0.05cm 1cm 0.815cm}, clip, height=0.85\columnwidth]{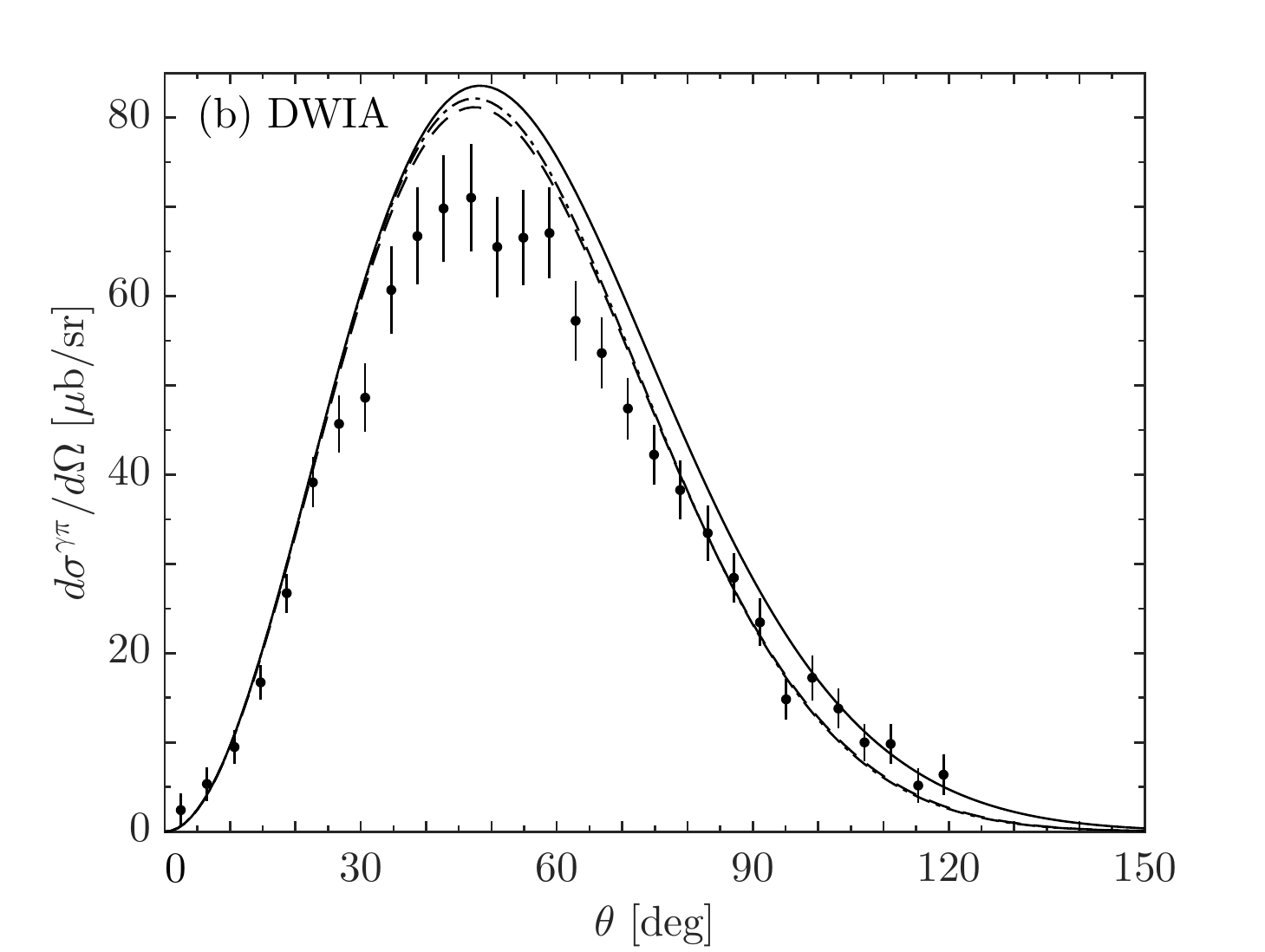}
        \caption{\label{f4C} Coherent $\pi^0$ photoproduction cross section on a $^{12}$C target for a 200~MeV incident photon. The calculations are shown within the (a) PWIA and (b) DWIA versions of the model presented in \Sec{sec:theory} using the densities displayed in Fig.~\ref{f2C}: the S\~ao Paulo (solid line), HO (dash-dotted line) and FB (dashed line) densities. Experimental data from Ref.~\cite{KRUSCHE2002287}$^{\ref{Note1}}$.}
\end{figure*}

The $\pi^0$-photoproduction cross section for a 200\,MeV photon impinging on a $^{12}$C target is displayed in Fig.\,\ref{f4C} as a function 
of the scattering angle $\theta$ in the $\pi$-$A$ center-of-momentum frame. The (a) PWIA and (b) DWIA results presented in \Sec{sec:theory} 
are compared to the experimental data of Ref.\cite{KRUSCHE2002287}\footnote{\label{Note1}Note that these data are extracted directly from 
Fig.~4 of Ref.~\cite{KRUSCHE2002287}. For some of these data points, the error bars are not visible in the figure as they are occulted 
by the dots representing the data. For these points, we have made the conservative hypothesis that the data uncertainty is as 
large as the size of the dots.}. We consider the three sets of model densities shown in \Fig{f2C} following the same line-type convention.

As is visible on panel (a) the PWIA cross sections are in a good agreement with the data without any adjustment of parameters.
Not only is the predicted shape similar to the experimental one, with a maximum around 50$^{\circ}$, but its magnitude is also
in agreement with experiment. Our calculations exhibit a rather small sensitivity to the choice of density: despite the clear differences 
noted in \Fig{f2C}, the corresponding $\pi^0$-photoproduction cross sections are very close to one another with the theoretical 
spread considerably smaller than the experimental error\,\cite{KRUSCHE2002287}. We note that the most compact spatial distribution,
namely, the one obtained with the S\~ao Paulo parametrization, leads to the cross section that extends to the largest angles. 
This is to be expected since the $\pi^0$-photoproduction amplitude listed in Eq.\eq{UPWIA} varies linearly with the Fourier transform 
of the nucleonic densities. The very similar HO and FB densities lead to nearly indistinguishable $\pi^0$-photoproduction cross 
sections.

When the reaction is calculated including final state interactions, [\Fig{f4C}(b)], the photoproduction cross section increases
by about 25\%. Although the agreement is not as good as in the PWIA, such a significant enhancement confirms that pionic distortions 
cannot be neglected\,\cite{NPA660,PLM98}. The energy dependence of the $\pi$-$A$ optical potential might be the cause for the observed
disagreement. Another possibility is the role played by higher-order effects, as emphasized in Ref.~\cite{Mil19}. Notably, pionic
distortions seem to influence only the magnitude of the cross section; there is no significant change in the shape of the theoretical 
cross sections. That the DWIA cross section is proportional to the PWIA one is reminiscent of what has been observed in 
Ref.~\cite{ZPA.328.195}, where the effect of distortion on coherent $\pi^0$ photoproduction has been analyzed for light targets 
and at energies near threshold, i.e., below 200\,MeV. At the peak of the cross section, we can note some minor changes in the
spreading of the theoretical predictions, suggesting that the choice of nuclear density is fairly insensitive to final-state interactions.

\begin{figure*}
	\includegraphics[trim={0.2cm 0.05cm 1.39cm 0.5cm}, clip, height=0.85\columnwidth]{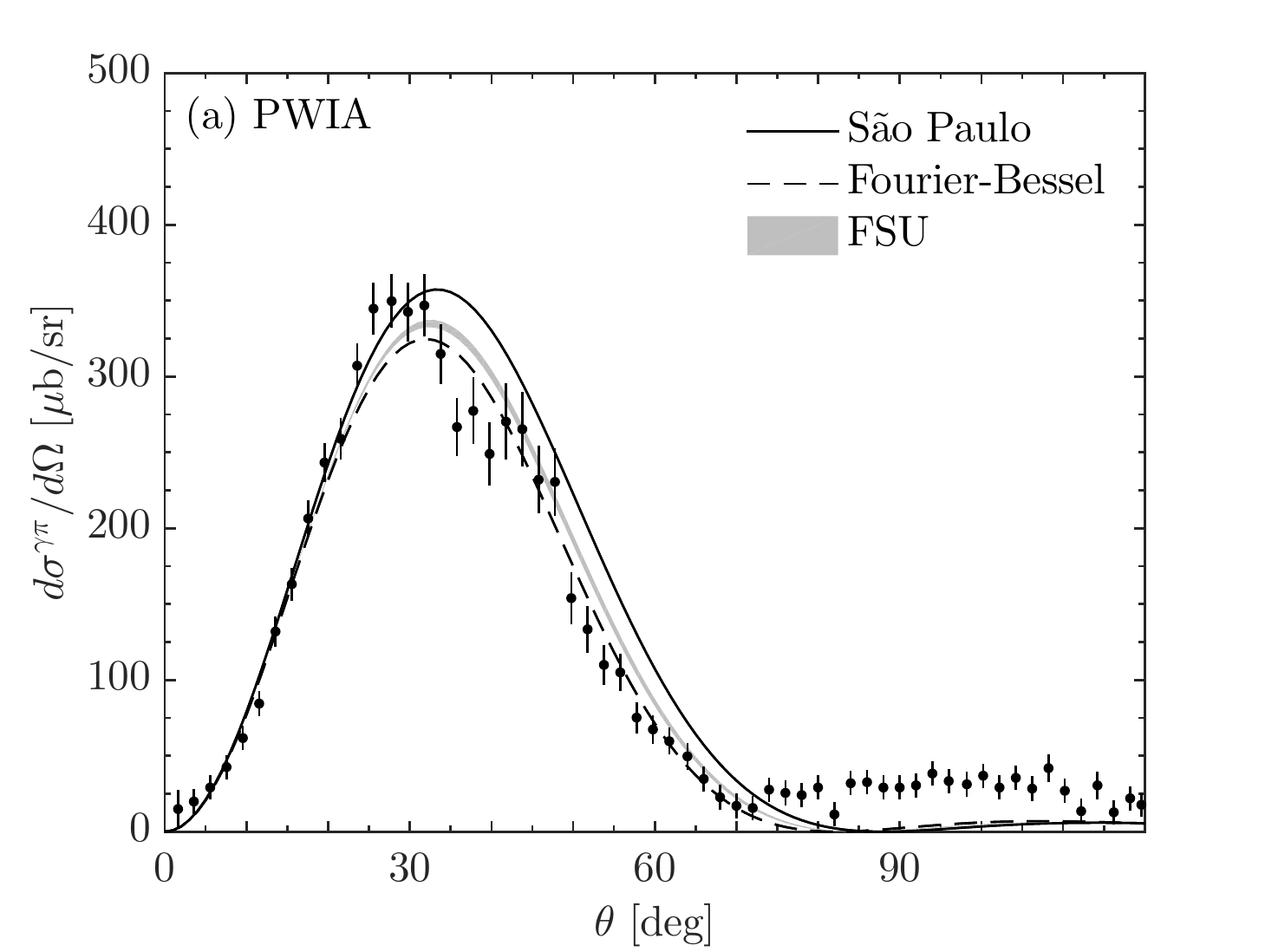}\includegraphics[trim={1.93cm 0.05cm 1cm 0.5cm}, clip, height=0.85\columnwidth]{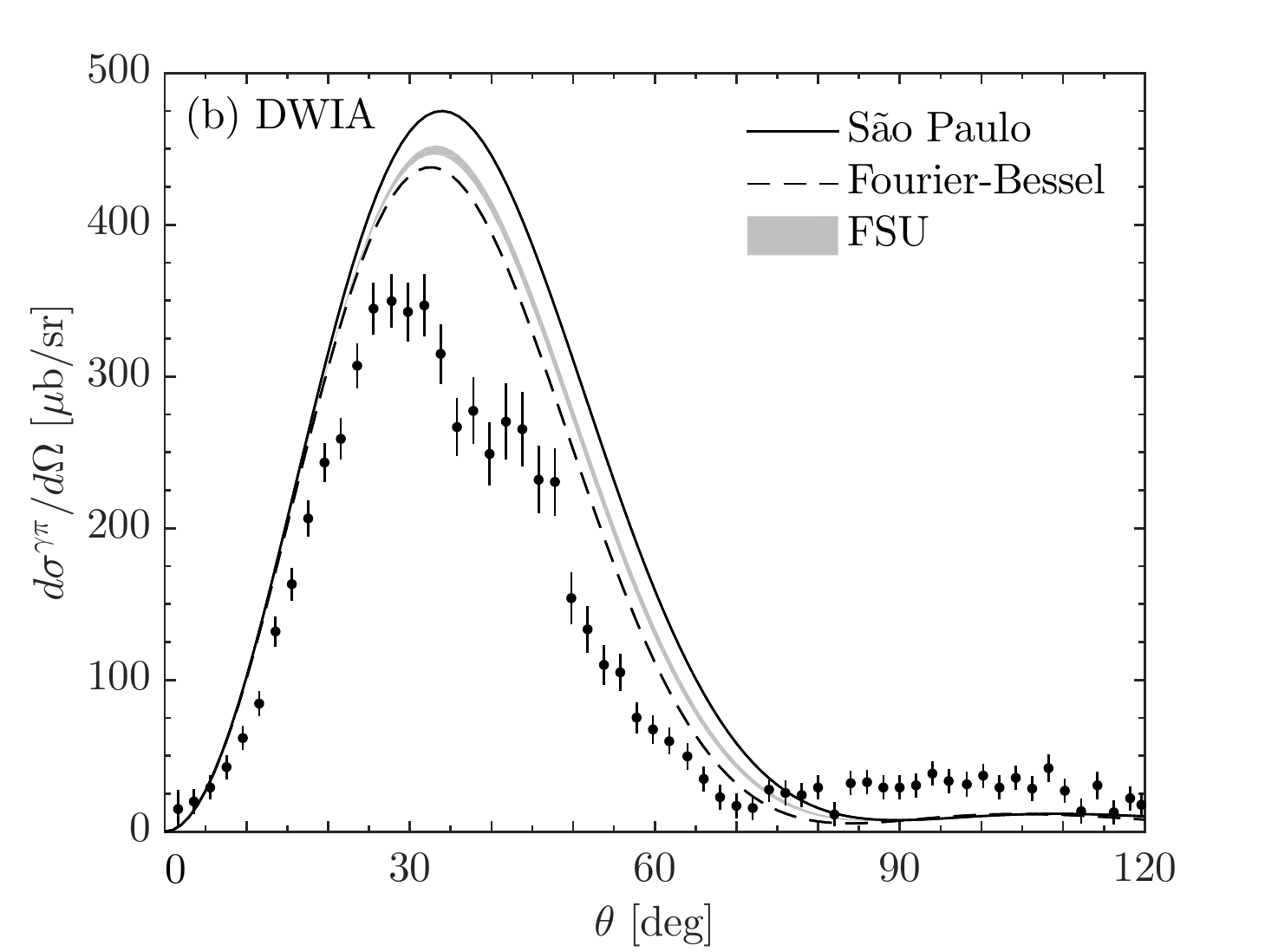}
	\caption{\label{f4Ca} Cross section for the $\pi^0$ photoproduction of a 200~MeV photon impinging on $^{40}$Ca computed at (a) the PWIA and (b) the DWIA.
	The calculations use the three densities presented in Fig.~\ref{f2Ca}: the S\~ao Paulo Fermi-Dirac density (solid line), Fourier-Bessel expansion of the charge density (dashed line) and the relativistic mean-field FSU calculation (grey band).
	Experimental data from Ref.~\cite{KRUSCHE2002287}.}
\end{figure*}

Figure~\ref{f4Ca} displays the $\pi^0$-photoproduction cross section for a 200\,MeV photon impinging on a $^{40}$Ca nucleus, 
(a) without and (b) with the inclusion of pionic distortions. The theoretical predictions are compared to the experimental data of 
Ref.\,\cite{KRUSCHE2002287}. On this heavier target, the nuclear form factor falls faster resulting in a cross section that is more 
forward focussed than in $^{12}$C, with the maximum located at $\theta\approx30^\circ$. As in the case of $^{12}$C, the shape 
and absolute magnitude are better reproduced by the PWIA calculations. We also note the appearance of a minimum in the data 
at about $70^\circ$ followed by a second broad maximum. Such behavior is only qualitatively reproduced by our calculations: 
without distortions, all three choices of density predict a node between $80^\circ$ and $90^\circ$ followed by a tiny bump. In turn, 
pionic distortions fill the node, resulting in a slight increase of the height of the second maximum.

These features---viz. more forward-focussed cross section and the emergence of a minimum---can be easily interpreted in the 
PWIA remembering that the $T$ matrix in Eq.\eq{UPWIA} is proportional to the Fourier transform of the nuclear densities defined
in Eq.\eq{FTdens}. When the target mass increases, so does the radial extension of the nuclear density, whose Fourier transform 
accordingly peaks at smaller momentum $q$. The zero observed in the PWIA calculations corresponds to the first zero of 
${\cal F}[\rho]$.

The differences observed in the cross sections computed with the three densities can also be easily interpreted following this 
line of thoughts. The spatial density that extends the farthest, i.e. the charge density described by a Fourier-Bessel expansion 
(see \Fig{f2Ca}), produces the cross section that peaks at the most forward angle. It is then followed by the FSU band, whose 
densities are slightly less extended. The S\~ao Paulo parametrization, which predicts a very narrow neutron density, in turn leads 
to the cross section that extends to the largest angles. Note however that these differences in the cross section remain small 
despite the large discrepancy observed between the nucleonic densities in \Fig{f2Ca}. The differences in the model are of the 
order of the experimental uncertainty, so it would be difficult to identify an optimal density choice solely based on these data.

As mentioned before, the distortion induced by the $\pi^0$-$A$ interaction significantly increases the theoretical predictions 
in the first peak in a way that seems independent of the nuclear density. Around the first maximum, we observe a uniform 
enhancement of the DWIA cross section by about 35\% relative to the PWIA prediction. At larger angles, the effect of the 
distortion effects is no longer linear. Disortions smear the PWIA curves in the region of their node and enhances the second 
maximum, however not enough to reach the experimental data. The disagreement with the data on $^{40}$Ca is stronger 
than for $^{12}$C. Part of the problem may again be due to the parameters of the MSU potential or to higher-order effects 
neglected in this IA approach\,\cite{Mil19}. Another major source of uncertainty is of experimental nature: in the case of $^{40}$Ca, 
it was difficult to isolate the purely coherent events from the incoherent ones\,\cite{Kru18}. Nevertheless, this test enables 
us to confirm that our model provides sensible results, that distortions play a non-negligible role, and, most importantly, that 
this observable depends little on the choice of the nuclear density.

\begin{figure*}
	\includegraphics[trim={0.1cm 0.05cm 1.39cm 0.8cm}, clip, height=0.85\columnwidth]{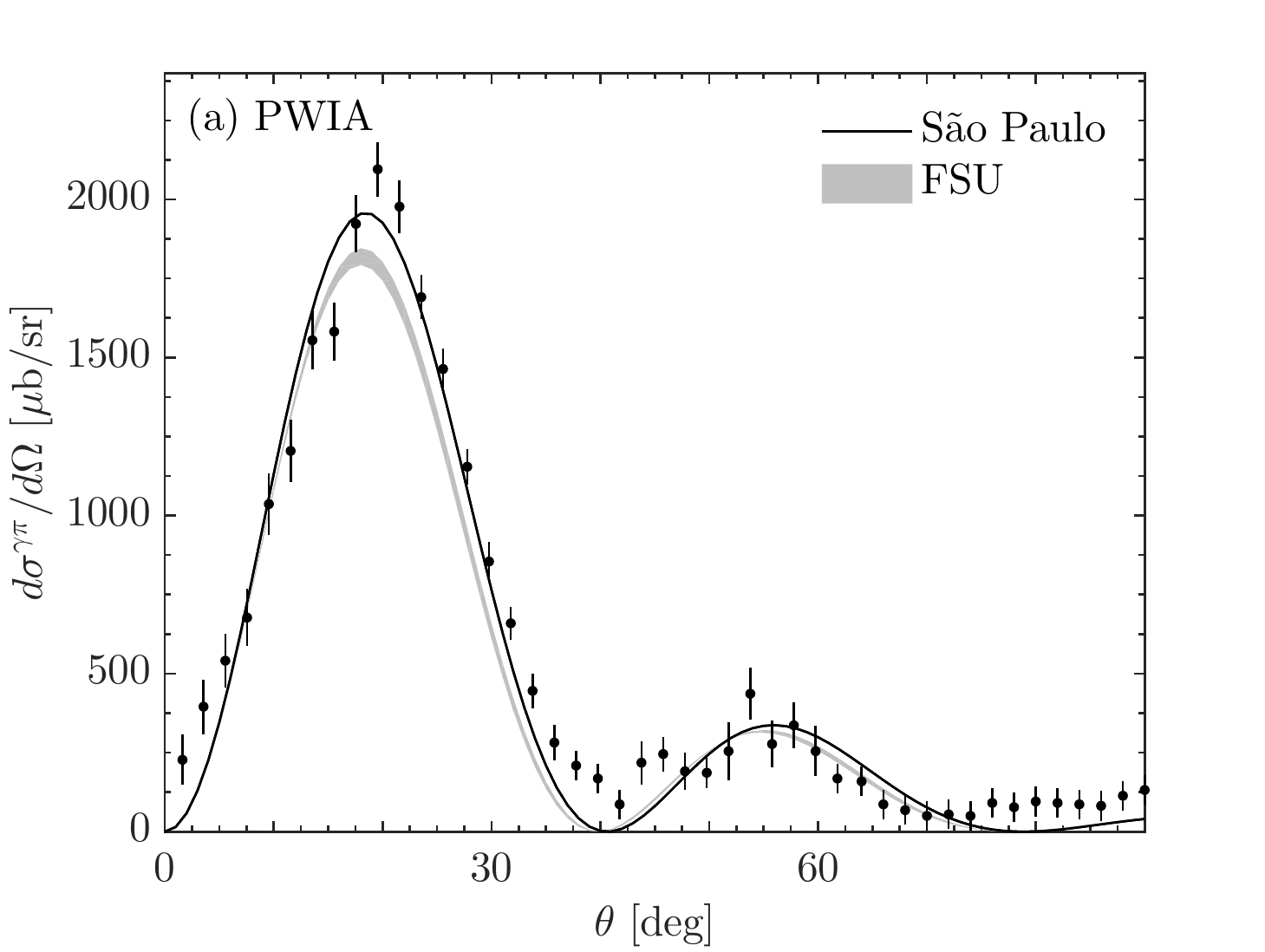}\includegraphics[trim={1.93cm 0.05cm 1cm 0.8cm}, clip, height=0.85\columnwidth]{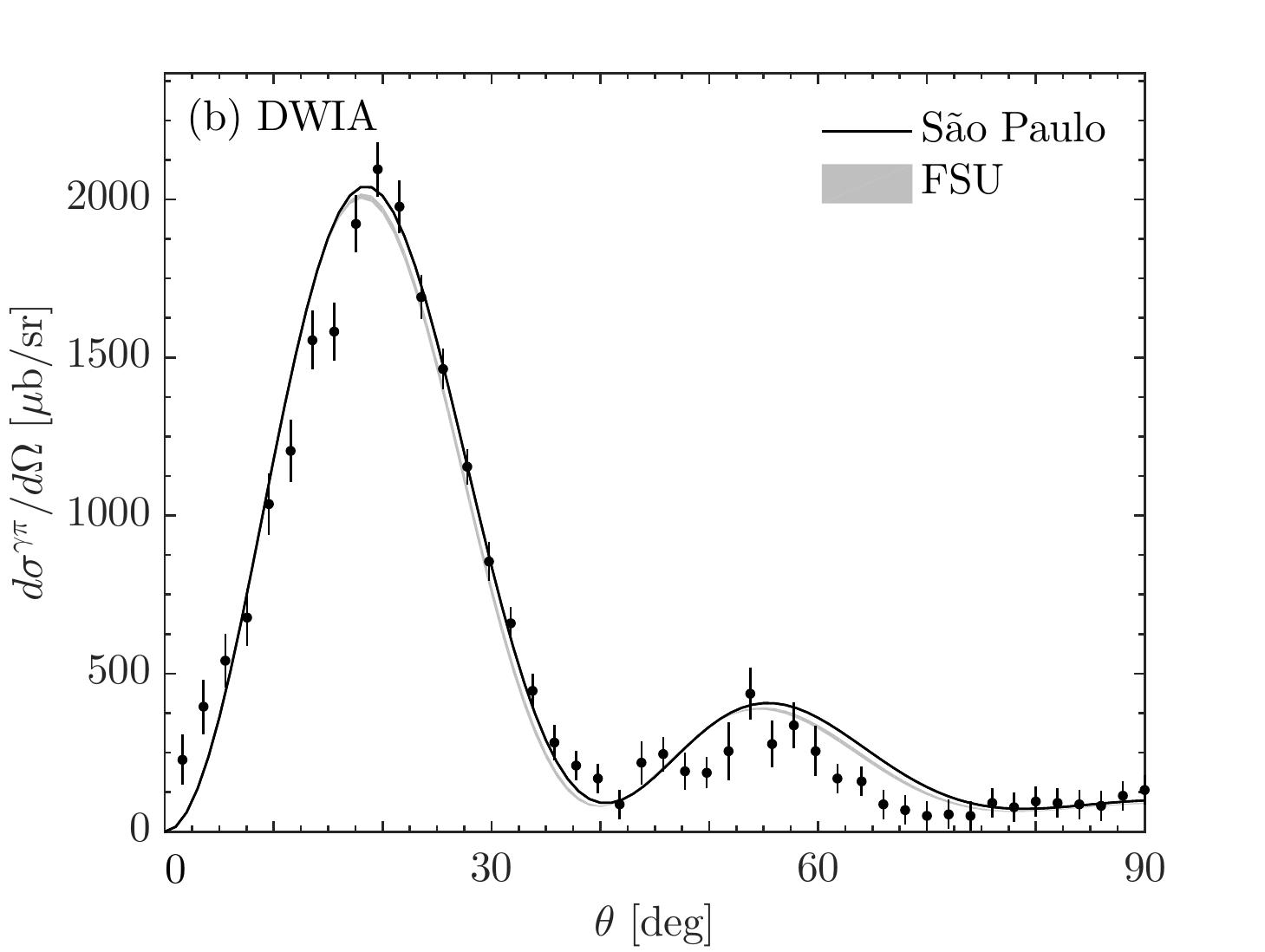}
	\caption{\label{f4Pb} 
	Cross section for the $\pi^0$ photoproduction cross section on $^{208}$Pb by a 200~MeV photon. The calculations performed at (a) the PWIA and (b) the DWIA are presented for the S\~ao Paulo density (solid line) and using the results of the FSU relativistic mean field calculations (grey band).
	Experimental data from Ref.~\cite{KRUSCHE2002287}.}
\end{figure*}

The $\pi^0$-photoproduction cross section on $^{208}$Pb induced by a 200\,MeV photon is presented in \Fig{f4Pb}. Both 
PWIA and DWIA calculations are compared against the experimental data of 
Ref.\,\cite{KRUSCHE2002287}\footnote{Unfortunately, we are unable to directly compare our results to the data of Tarbert \etal\ \cite{PhysRevLett112_242502} because of an inconsistency in the presentation of this more recent measurement on $^{208}$Pb, see Ref.~\cite{Mil19}. However, we were told \cite{Kru18} that with a proper scaling, they are in excellent agreement with the previous data of Krusche \etal\ \cite{KRUSCHE2002287}. Accordingly, we infer that this new set of data does not change the conclusion of our study.} We consider the two density choices presented in \Fig{f2Pb}. 

The changes observed for the lead target follow closely the insights developed earlier. The cross section on such a heavy nucleus 
peaks at even more forward angles, with the first maximum at $\theta\approx18^\circ$ and a first minimum at about $40^\circ$; a 
second maximum is now clearly visible at $55^\circ$ followed by a second minimum and the shadow of a subsequent third maximum 
beyond $80^\circ$. Given that these features are encoded in the nucleonic form factors, they are already evident at the PWIA level.
Indeed, the first and second experimental peaks\,\cite{KRUSCHE2002287} are well reproduced both in location and magnitude.
The first minimum is also well reproduced, although the theoretical cross section reaches zero where the Fourier transform of the 
nucleonic density vanishes. Once again, we observe a rather small difference between the model densities that is easy to 
understand: given that the S\~ao Paulo densities decay faster than the FSU ones, the associated cross section is shifted to 
slightly larger angles (see \Fig{f2Pb}).

As was the case for $^{12}$C and $^{40}$Ca, introducing pionic final-state interactions leads to an increase in the cross section in the first peak.
However, in this case, accounting for distortion effects \emph{improves} the description of the experimental data.
The magnitude of the first peak predicted by the DWIA calculations is in excellent agreement with the experiment.
Moreover, the  smearing of the first node is such that our theoretical cross sections now fall directly on the data.
It is interesting to note that the differences observed at the PWIA between the two sets of nucleonic densities is reduced when distortions are included, so that both calculations agree nearly perfectly with the data.
Given that these calculations  have been performed without any adjustment of parameters, it is gratifying to see such an excellent description of the experimental data\,\cite{KRUSCHE2002287}.
Relative to the lighter targets, the excellent agreement observed here may be  due to the fact that the IA may be better suited for heavy targets. 

Since both sets of nuclear densities provide a nearly perfect description of the experimental data, it is clear that no reliable nuclear
structure information can be obtained solely from this experiment. This result has a profound implication on the use of 
coherent $\pi^0$ photoproduction to infer the neutron skin thickness of heavy nuclei. As shown in \Fig{f4Pb}(b), using very different 
nucleonic densities---from the purely phenomenological S\~ao Paulo parametrization to a range of relativistic mean-field models---we 
observe no significant differences in our predictions for $^{208}$Pb. This finding is in stark disagreement with the result reported in 
Ref.\,\cite{PhysRevLett112_242502}, that suggests that the neutron skin thickness of $^{208}$Pb can be extracted with an astounding
precision of $R_{\rm skin}^{208}\!=\!0.15\pm0.03(\mathrm{stat.})^{+0.01}_{-0.03}(\mathrm{sys.})$\,fm. In contrast, our analysis indicates 
that this photoproduction reaction cannot distinguish between theoretical models for which the neutron skin thickness in $^{208}$Pb 
differs by at least 0.2\,fm. We note that this interval is likely a lower bound since calculations performed with values outside the adopted
range may still provide an equally good description of the experimental data. Beyond this systematic uncertainty, one should 
also take into account various higher-order effects that are not included within the framework of the impulse approximation. For example, 
 Miller shows that although small, these effects can result in an additional theoretical uncertainty of up to 6\%, which can falsify the inferred 
 neutron skin thickness by up to 50\%\,\cite{Mil19}.

Our analysis indicates that the coherent photoproduction amplitude is largely isoscalar. That is, the cross section is sensitive to the whole 
(neutron\,+\,proton) density and bears little dependence on the differences in densities. Accordingly, this reaction is not suitable to 
constrain any isovector observable, such as the neutron skin thickness. Perhaps by comparing $\pi^0$-photoproduction cross sections 
along a long chain of stable isotopes may provide a better constraint on $R_{\rm skin}$. We explore this possibility in the following section.

\section{\label{sec:Sn}Predictions for $\pi^0$ photoproduction on  $^{116,\,124}\rm Sn$}

Tin exhibits the longest chain of stable isotopes in the nuclear chart, so it would be interesting to examine how the neutron skin thickness evolves with
increasing neutron number.  Inspired by the precision claimed in Ref.\,\cite{PhysRevLett112_242502}, an experimental campaign has been started at 
MAMI, the Mainz Microtron of the Johannes Gutenberg University Mainz \cite{MAMI}, to measure the coherent $\pi^0$ photoproduction on three stable 
isotopes $^{116}$Sn, $^{120}$Sn and $^{124}$Sn. In this section we apply our theoretical framework to the lightest and heaviest of these isotopes to
(a) assess how changes in the nuclear density, specially in the neutron skin thickness, affect the $\pi^0$ photoproduction cross sections for each of 
these two nuclei and (b) to estimate the experimental precision required to resolve the expected differences in the neutron skin thickness of these two
nuclei. The measurements have been carried out in a range of incoming photon energies between 140\,MeV and 300\,MeV. Since the extraction of 
coherent events is optimized within the 180--190\,MeV energy bin, we focus on this particular energy range, which is compatible with the MSU potential 
used to describe the final $\pi^0$-nucleus interaction \cite{PhysRevC25_952}.

\begin{figure}
	\includegraphics[trim={0 1.2cm 1cm 0cm}, clip, width=\columnwidth]{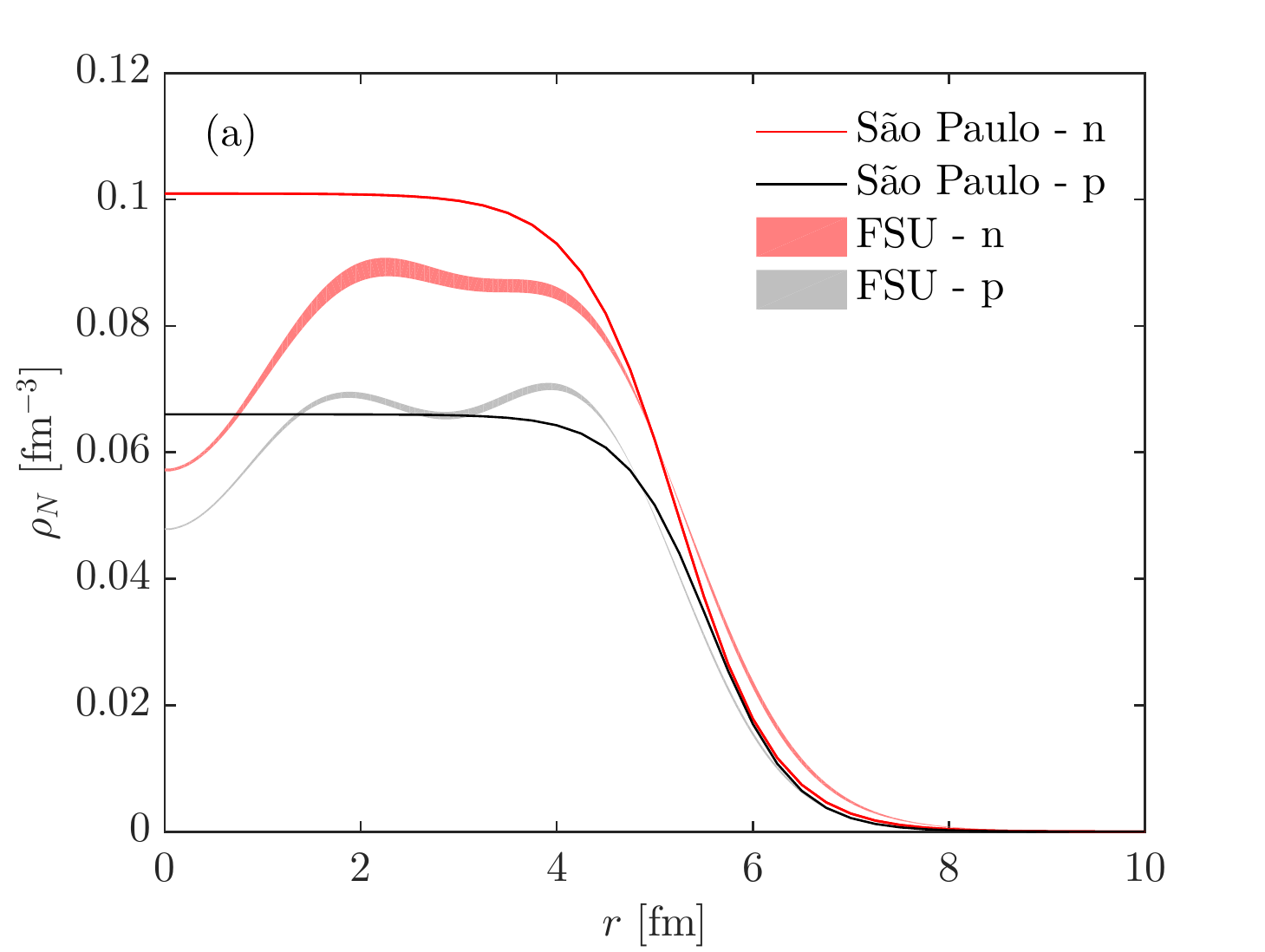}
	\includegraphics[trim={0 0.05cm 1cm 0cm}, clip, width=\columnwidth]{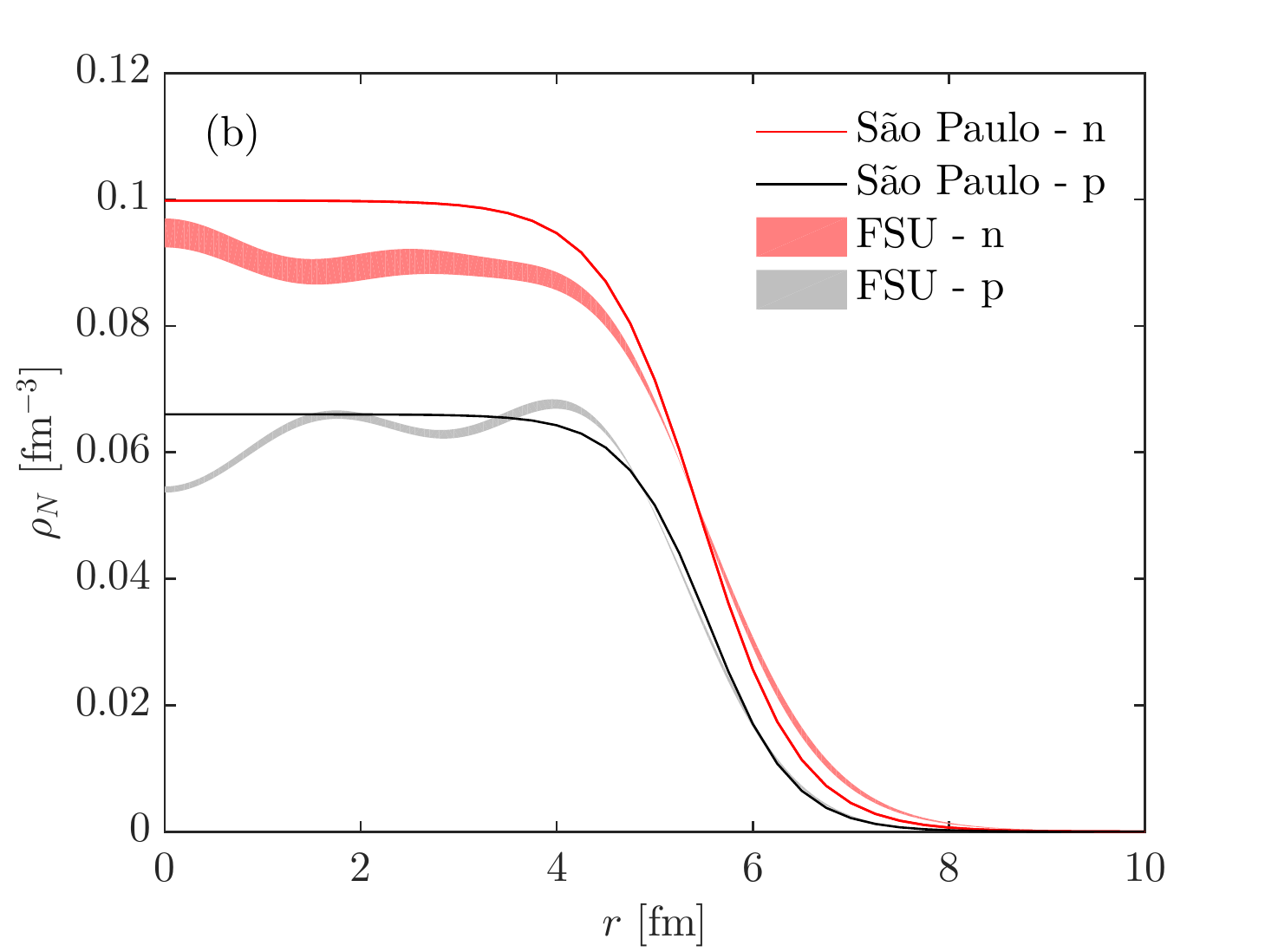}
	\caption{\label{f5} Neutron (red) and proton (black/grey) densities for (a) $^{116}$Sn and (b) $^{124}$Sn. S\~ao Paulo (solid lines) \cite{PhysRevC.66.014610} and FSU (bands) \cite{PhysRevLett.95.122501} densities are displayed.}
\end{figure}

As in the case of $^{208}$Pb, we consider nucleonic densities obtained from both the phenomenological S\~ao Paulo\,\cite{PhysRevC.66.014610} 
and FSU\,\cite{PhysRevLett.95.122501} parametrizations. As done earlier, we consider the same set of relativistic mean field models to produce a 
set of physically meaningful densities that will generate a broad range of neutron skin thickness. These density profiles are displayed in \Fig{f5} for
(a) $^{116}$Sn and (b) $^{124}$Sn. The S\~ao Paulo parametrization is shown with solid lines and the range of FSU predictions shown as bands; 
the black/grey lines correspond to proton densities and the red ones to neutron densities. The S\~ao Paulo potential predicts a profile that decays 
faster with $r$ than the mean-field calculations, specially in the case of the neutron densities. We note that the interior depression in $^{116}$Sn,
the so-called ``nuclear bubble"\,\cite{Todd-Rutel:2004llz,Grasso:2009zza}, is due to an empty $3s^{1/2}$ neutron orbital that is only 0.4\,MeV below 
the fully occupied $2d^{3/2}$ orbital. We expect that the bubble will be partially filled by the inclusion of pairing correlations which are absent from 
our model. In the case of $^{124}$Sn, both of these neutron orbitals are filled so the the bubble disappears. The neutron skin thickness deduced 
from these densities can be found in Table~\ref{tab2}. Based on the S\~ao Paulo density profiles shown in \Fig{f5}, the S\~ao Paulo potential predicts 
a negative neutron skin thickness for $^{116}$Sn and a very thin one for $^{124}$Sn. In contrast, the FSU parametrization predicts significantly 
thicker neutron skins and illustrates the clear impact of adding 8 more neutrons in going from $^{116}$Sn to $^{124}$Sn.

\begin{figure*}[htb]
	\includegraphics[trim={0.095cm 0.4cm 1.39cm 0.46cm}, clip, height=0.85\columnwidth]{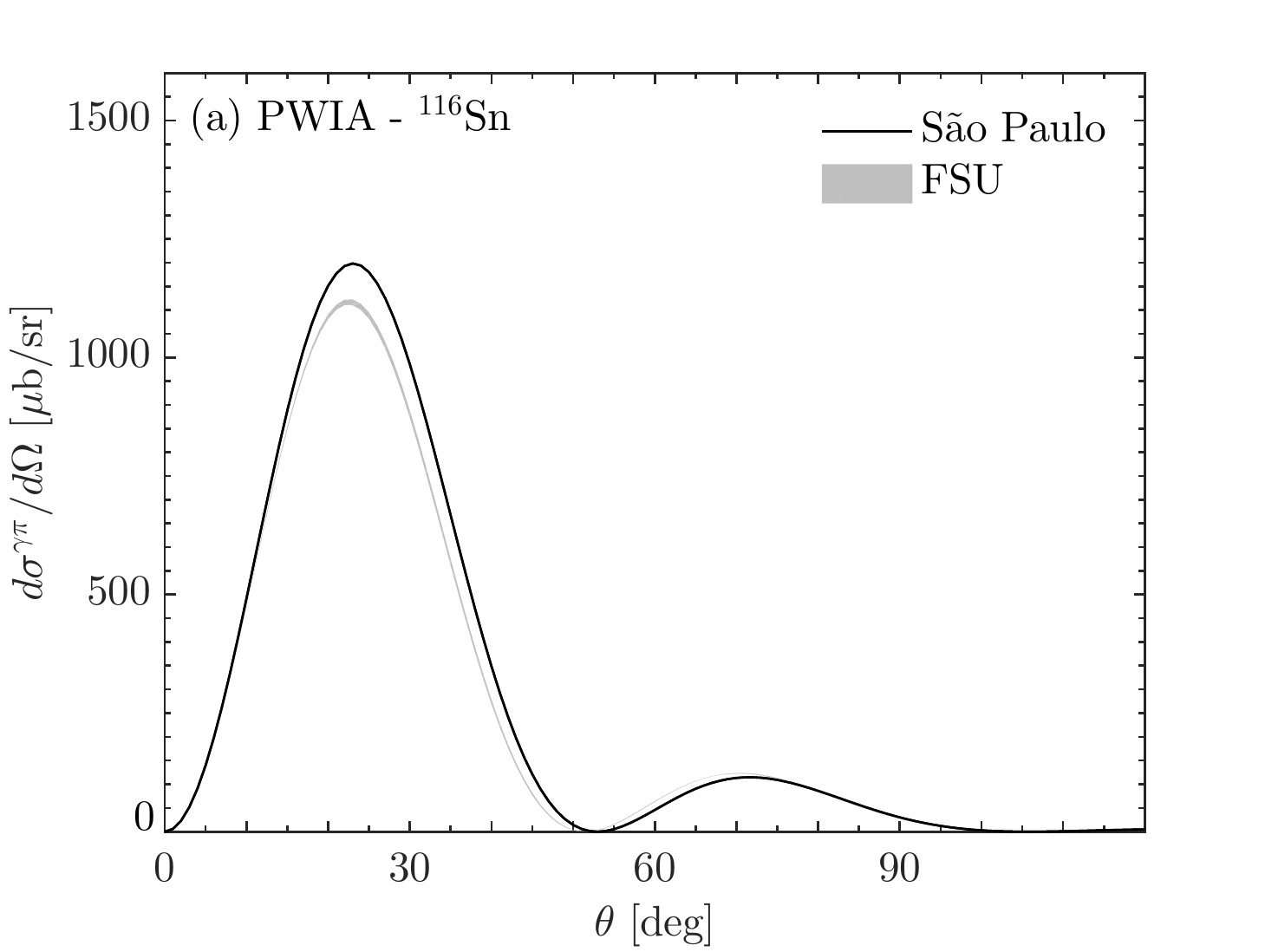}\includegraphics[trim={1.93cm 0.4cm 1cm 0.46cm}, clip, height=0.85\columnwidth]{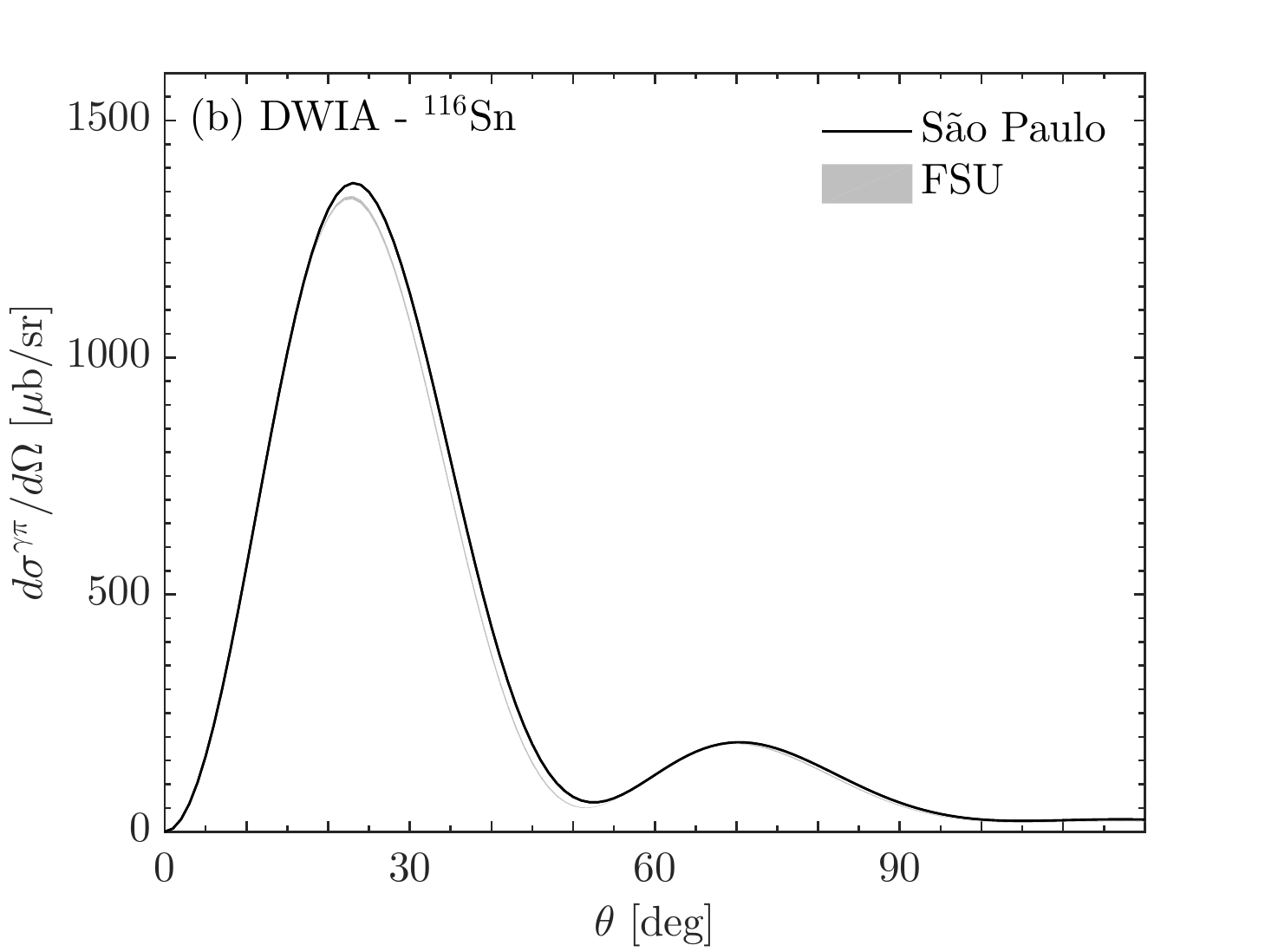}
		
	\vspace{-0.75cm}
	\includegraphics[trim={0.1cm 0.05cm 1.39cm 0.81cm}, clip, height=0.8473\columnwidth]{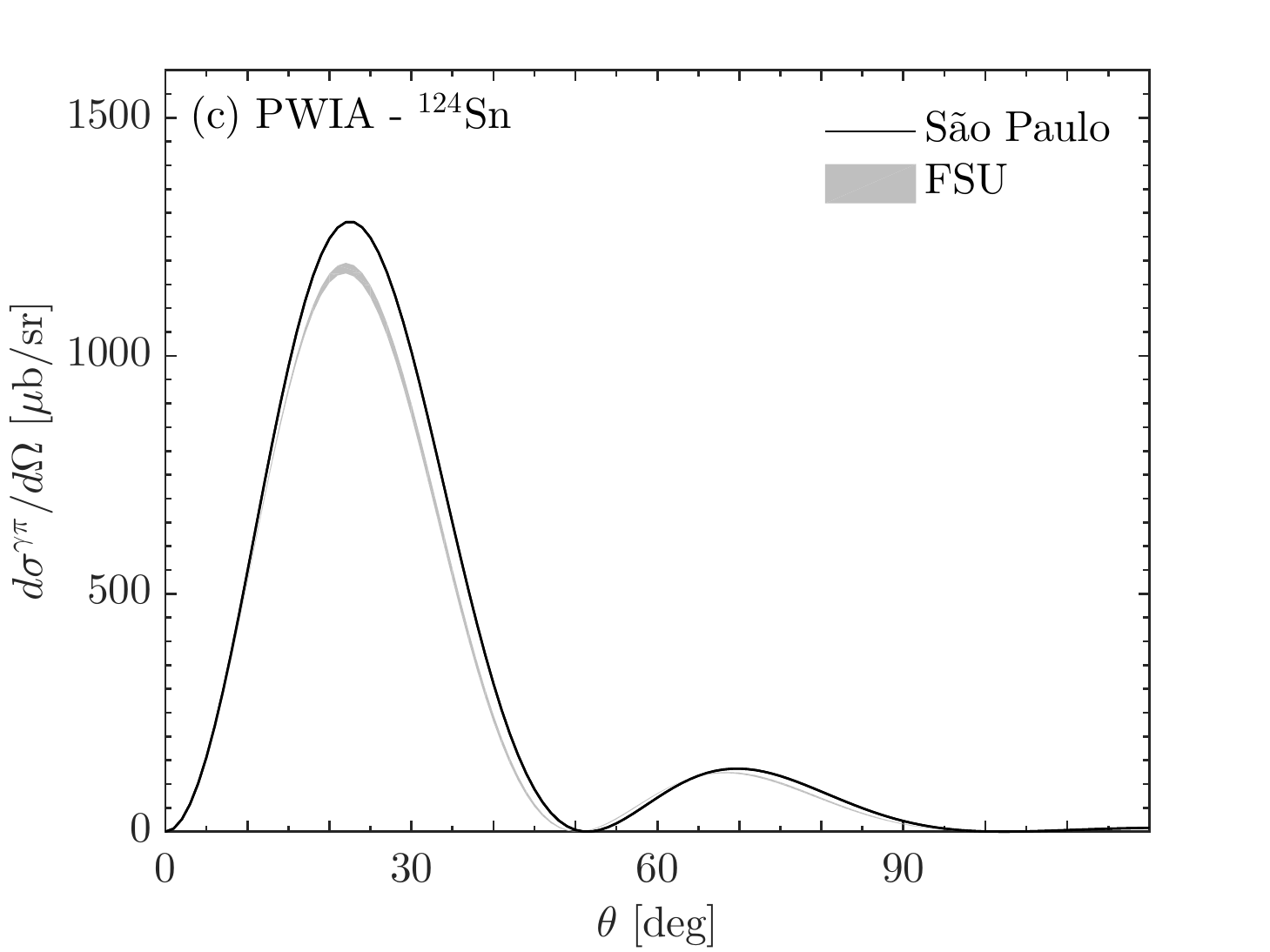}\includegraphics[trim={1.93cm 0.05cm 1cm 0.845cm}, clip, height=0.8473\columnwidth]{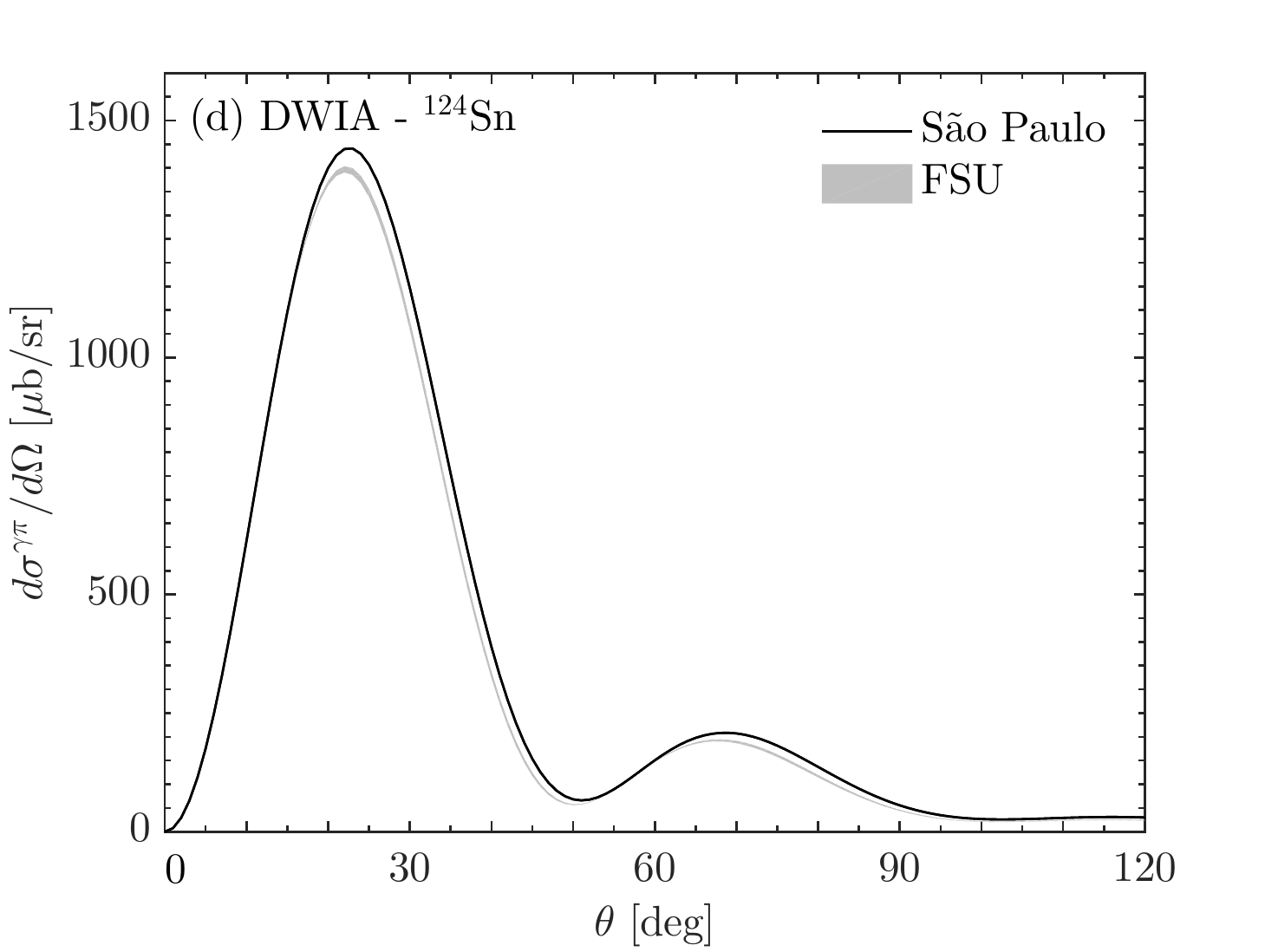}
	
	\caption{\label{f6} Coherent $\pi^0$ photoproduction cross section for a 200~MeV photon impinging on $^{116}$Sn (top panels) and $^{124}$Sn (bottom panels).
The calculations are performed at the PWIA [(a) and (c)] and at the DWIA [(b) and (d)] using the density profiles shown in Fig.~\ref{f5}: the S\~ao Paulo parametrization (solid line) and the prediction of the FSU relativistic mean field model (grey band).}
\end{figure*}

\begin{table}[b]
	\caption{\label{tab2} Neutron skin thicknesses for $^{116}$Sn and $^{124}$Sn.}
	\renewcommand{\arraystretch}{1}
	\centering
	\makebox[\columnwidth]{\begin{tabular}{c | c  c}\hline\hline
			Model & S\~ao Paulo & FSU range \\
			\hline
			$^{116}$Sn & -0.035~fm & [0.104, 0.170]~fm  \\
			$^{124}$Sn & 0.013~fm & [0.188, 0.284]~fm\\ \hline\hline
	\end{tabular}}
\end{table}

Our calculated cross sections are presented in Fig.~\ref{f6}. The top panels correspond to the $^{116}$Sn target with calculations 
performed in the (a) plane-wave and (b) distorted-wave impulse approximation; the bottom panels show the corresponding results 
for $^{124}$Sn. As expected, the main features of these results can be readily understood from the analysis presented in the previous 
section. First, with a first maximum between $22^\circ$ and $23^\circ$ and a first minimum at about $50^\circ$, the angular distributions 
on the tin isotopes, while more forward focussed than on $^{40}$Ca, spread to larger angles than on $^{208}$Pb. We note that if
the nuclear density could be faithfully approximated by a \emph{symmetrized} Fermi-Dirac shape---which is practically identical to 
the conventional Fermi-Dirac shape---the associated form factor is known analytically\,\cite{Piekarewicz:2016vbn}. Such an analytic
approach encapsulates the main features displayed by the cross section, namely, diffractive oscillations controlled by the half-density 
radius modulated by an exponential falloff controlled by the surface thickness. Second, whereas \Fig{f5} suggests significant differences
in the predictions of the nuclear-structure models for the proton and neutron densities, the sensitivity of the cross section to the choice of density profile is rather 
small. The minor changes observed between the cross sections computed with the S\~ao Paulo and FSU densities can once more be 
traced back to their Fourier transform: the S\~ao Paulo parametrization, predicting a more compact density profile, leads to a slightly 
more extended $\pi^0$-photoproduction cross section than the FSU calculations. That is, despite the significant differences in the
predictions for the neutron skin thickness (see Table~\ref{tab2}) the minimum in the cross section suggested by the S\~ao Paulo 
parametrization is shifted to a slightly larger angle relative to the FSU predictions. Third, the inclusion of final-state interactions 
[panels (b) and (d)] has a significant effect: it increases the cross section at all angles although not uniformly; for example, the first 
and second minima are filled whereas the increase in the first peak is about 10\%. Moreover, as for $^{208}$Pb, we observe a clear reduction of the 
relative difference between the calculations performed with the S\~ao Paulo and the FSU densities. This is especially noticeable 
in \Fig{f6} (b) for the case of $^{116}$Sn, for which the pionic distortions wash away most of the differences. The third point 
confirms that once the $\pi^0$-nucleus interaction is accounted for in the outgoing channel, all densities---despite their
significant differences---provide very similar cross sections. In the particular case of the more realistic FSU densities, which cover 
a range of neutron skin thicknesses of about 0.07\,fm for $^{116}$Sn and 0.1\,fm for $^{124}$Sn, the differences in our 
DWIA cross sections do not even reach 1.5\% at the first maximum for either isotope. These results suggest that even if 
the experiments could reach a 1.5\% precision, the neutron skin thickness could not be determined to better than 
0.07~fm for $^{116}$Sn and 0.1~fm for $^{124}$Sn.

Because of the small overall difference in size between both isotopes, we observe a slight shift towards forward angles of the 
$\pi^0$-photoproduction cross section from $^{116}$Sn to $^{120}$Sn. This isotopic shift varies slightly with the choice of the 
density. However, as for the aforementioned difference in magnitude, the model dependence is reduced at the DWIA level. To 
resolve such a minor shift in order to get some information on the difference in neutron skin thickness between both targets, 
an experimental precision of at least $1.5^\circ$ is required. Perhaps a combination of these cross sections, such as a relative 
difference, might enable us to infer a more reliable estimate of the change in neutron skin thickness between the tin isotopes.
However, as already seen in \Sec{sec:CCaPb}, under the present experimental conditions, it is unrealistic to use this sole reaction 
observable to infer the absolute neutron skin thickness of these nuclei. Ultimately, it is the fact that the $\pi^0$-photoproduction 
reaction is mostly sensitive to the \emph{isoscalar} density that makes the extraction of the isovector neutron skin thickness so 
challenging.

\section{\label{sec:conclusion}Conclusion}

Coherent $\pi^0$ photoproduction has been suggested as an accurate probe of the nuclear shape with sufficient sensitivity 
to determine the thickness of the neutron skin\,\cite{PhysRevLett112_242502}. In this work, we test this assertion by performing
calculations of the $\pi^0$-photoproduction cross section using a distorted wave impulse approximation framework to compare
against existing and future experimental data. We consider $^{12}$C, $^{40}$Ca, $^{116,124}$Sn, and $^{208}$Pb as targets
and use a variety of nucleonic densities to test the sensitivity of the cross section to the neutron skin thickness. 

We observe a general good agreement with the data of Ref.~\cite{KRUSCHE2002287}.
For the light targets $^{12}$C and $^{40}$Ca 
we reproduce the shape and order of magnitude of the experimental cross section. In the case of $^{208}$Pb, the agreement is nearly 
perfect without resorting to any parameter fitting, probably because the impulse approximation and a mean-field description of nuclear
densities are better suited for heavier targets. As observed in Ref.~\cite{NPA660,KRUSCHE2002287}, the effect of final-state interactions 
is significant so it cannot be ignored. Pionic distortions lead to an increase of the cross section, which is largest on the light targets.
Moreover, final-state interaction are largely insensitive to the choice of nuclear density. In general---and particularly for the heavier 
targets---such an effects tends to \emph{reduce} the sensitivity of the $\pi^0$-photoproduction cross section to the nuclear density.

The major result of this work is that the photoproduction of a neutral pion on a nucleus is largely insensitive to the nuclear density.
In the case of $^{208}$Pb, using densities generated from a relativistic mean-field model\,\cite{PhysRevLett.95.122501} produces 
a cross section that is practically indistinguishable from the one obtained using a simple Fermi-Dirac parametrization that predicts
a very small neutron skin\,\cite{PhysRevC.66.014610}; such a small neutron skin has now been ruled out by experiment\,\cite{PREXII}.
In essence, given that the photoproduction cross section is dominated by the isoscalar ($\rho_{n}\!+\!\rho_{p}$) density, it is difficult 
to extract an isovector observable---such as the neutron skin thickness---with enough precision to provide meaningful constraints
on the density dependence of the symmetry energy. Indeed, our analysis shows that for the fairly wide range of neutron skin thickness 
adopted in this work, {\sl i.e.,} $R_{\rm skin}^{208}\!\approx\!(0.1\!-\!0.3)$\,fm, all models reproduce the experimental data of 
Ref.~\cite{KRUSCHE2002287} equally well, without any adjustment of model parameters. 

Our results challenge the main conclusion derived from a similar experiment that suggests that the neutron skin thickness of 
$^{208}$Pb can be determined with a total (statistical\,+\,systematic) uncertainty of only 
$\sim\!0.04$\,fm\,\cite{PhysRevLett112_242502}. Given that the $\pi^{0}$-photoproduction reaction is unable to provide a 
reliable estimate of the neutron skin thickness, the method cannot shed any light on the discrepancy observed by the various 
estimates of $R_{\rm skin}^{208}$ reported in the literature\,\cite{Tsa12,Thi19}.

The extension of this study to $^{116}$Sn and $^{124}$Sn---for which the $\pi^0$ photoproduction has already been measured 
at Mainz---leads to similar conclusions, confirming that this reaction is largely insensitive to the the thickness of the neutron skin.
Hence, these measurements alone are not sufficient to estimate accurately the evolution of the neutron skin thickness along that 
isotopic chain as initially hoped. A combination of different measurements, like a relative difference of cross sections measured 
on two extreme isotopes may help validate the idea behind the method.

\begin{acknowledgments}
We dedicate this article to the memory of Prof.\ Dr.\ Bernd Krusche, who passed away on 1 June 2022.
This work was supported by the PRISMA+ (Precision Physics, Fundamental Interactions and Structure of Matter) Cluster 
of Excellence and the Deutsche Forschungsgemeinschaft (DFG, German Research Foundation) under the 
Projekt-ID 204404729 (SFB 1044). J.P. acknowledges support by the U.S. Department of Energy Office of Science, Office 
of Nuclear Physics under Award DE-FG02-92ER40750.
\end{acknowledgments}


\begin{thebibliography}{38}%
\makeatletter
\providecommand \@ifxundefined [1]{%
 \@ifx{#1\undefined}
}%
\providecommand \@ifnum [1]{%
 \ifnum #1\expandafter \@firstoftwo
 \else \expandafter \@secondoftwo
 \fi
}%
\providecommand \@ifx [1]{%
 \ifx #1\expandafter \@firstoftwo
 \else \expandafter \@secondoftwo
 \fi
}%
\providecommand \natexlab [1]{#1}%
\providecommand \enquote  [1]{``#1''}%
\providecommand \bibnamefont  [1]{#1}%
\providecommand \bibfnamefont [1]{#1}%
\providecommand \citenamefont [1]{#1}%
\providecommand \href@noop [0]{\@secondoftwo}%
\providecommand \href [0]{\begingroup \@sanitize@url \@href}%
\providecommand \@href[1]{\@@startlink{#1}\@@href}%
\providecommand \@@href[1]{\endgroup#1\@@endlink}%
\providecommand \@sanitize@url [0]{\catcode `\\12\catcode `\$12\catcode
  `\&12\catcode `\#12\catcode `\^12\catcode `\_12\catcode `\%12\relax}%
\providecommand \@@startlink[1]{}%
\providecommand \@@endlink[0]{}%
\providecommand \url  [0]{\begingroup\@sanitize@url \@url }%
\providecommand \@url [1]{\endgroup\@href {#1}{\urlprefix }}%
\providecommand \urlprefix  [0]{URL }%
\providecommand \Eprint [0]{\href }%
\providecommand \doibase [0]{http://dx.doi.org/}%
\providecommand \selectlanguage [0]{\@gobble}%
\providecommand \bibinfo  [0]{\@secondoftwo}%
\providecommand \bibfield  [0]{\@secondoftwo}%
\providecommand \translation [1]{[#1]}%
\providecommand \BibitemOpen [0]{}%
\providecommand \bibitemStop [0]{}%
\providecommand \bibitemNoStop [0]{.\EOS\space}%
\providecommand \EOS [0]{\spacefactor3000\relax}%
\providecommand \BibitemShut  [1]{\csname bibitem#1\endcsname}%
\let\auto@bib@innerbib\@empty
\bibitem [{\citenamefont {Tsang}\ \emph {et~al.}(2012)\citenamefont {Tsang},
  \citenamefont {Stone}, \citenamefont {Camera}, \citenamefont {Danielewicz},
  \citenamefont {Gandolfi}, \citenamefont {Hebeler}, \citenamefont {Horowitz},
  \citenamefont {Lee}, \citenamefont {Lynch}, \citenamefont {Kohley},
  \citenamefont {Lemmon}, \citenamefont {M\"oller}, \citenamefont {Murakami},
  \citenamefont {Riordan}, \citenamefont {Roca-Maza}, \citenamefont
  {Sammarruca}, \citenamefont {Steiner}, \citenamefont {Vida\~na},\ and\
  \citenamefont {Yennello}}]{Tsa12}%
  \BibitemOpen
  \bibfield  {author} {\bibinfo {author} {\bibfnamefont {M.~B.}\ \bibnamefont
  {Tsang}}, \bibinfo {author} {\bibfnamefont {J.~R.}\ \bibnamefont {Stone}},
  \bibinfo {author} {\bibfnamefont {F.}~\bibnamefont {Camera}}, \bibinfo
  {author} {\bibfnamefont {P.}~\bibnamefont {Danielewicz}}, \bibinfo {author}
  {\bibfnamefont {S.}~\bibnamefont {Gandolfi}}, \bibinfo {author}
  {\bibfnamefont {K.}~\bibnamefont {Hebeler}}, \bibinfo {author} {\bibfnamefont
  {C.~J.}\ \bibnamefont {Horowitz}}, \bibinfo {author} {\bibfnamefont
  {J.}~\bibnamefont {Lee}}, \bibinfo {author} {\bibfnamefont {W.~G.}\
  \bibnamefont {Lynch}}, \bibinfo {author} {\bibfnamefont {Z.}~\bibnamefont
  {Kohley}}, \bibinfo {author} {\bibfnamefont {R.}~\bibnamefont {Lemmon}},
  \bibinfo {author} {\bibfnamefont {P.}~\bibnamefont {M\"oller}}, \bibinfo
  {author} {\bibfnamefont {T.}~\bibnamefont {Murakami}}, \bibinfo {author}
  {\bibfnamefont {S.}~\bibnamefont {Riordan}}, \bibinfo {author} {\bibfnamefont
  {X.}~\bibnamefont {Roca-Maza}}, \bibinfo {author} {\bibfnamefont
  {F.}~\bibnamefont {Sammarruca}}, \bibinfo {author} {\bibfnamefont {A.~W.}\
  \bibnamefont {Steiner}}, \bibinfo {author} {\bibfnamefont {I.}~\bibnamefont
  {Vida\~na}}, \ and\ \bibinfo {author} {\bibfnamefont {S.~J.}\ \bibnamefont
  {Yennello}},\ }\href {\doibase 10.1103/PhysRevC.86.015803} {\bibfield
  {journal} {\bibinfo  {journal} {Phys. Rev. C}\ }\textbf {\bibinfo {volume}
  {86}},\ \bibinfo {pages} {015803} (\bibinfo {year} {2012})}\BibitemShut
  {NoStop}%
\bibitem [{\citenamefont {Horowitz}\ \emph {et~al.}(2014)\citenamefont
  {Horowitz}, \citenamefont {Brown}, \citenamefont {Kim}, \citenamefont
  {Lynch}, \citenamefont {Michaels}, \citenamefont {Ono}, \citenamefont
  {Piekarewicz}, \citenamefont {Tsang},\ and\ \citenamefont {Wolter}}]{Hor14}%
  \BibitemOpen
  \bibfield  {author} {\bibinfo {author} {\bibfnamefont {C.~J.}\ \bibnamefont
  {Horowitz}}, \bibinfo {author} {\bibfnamefont {E.~F.}\ \bibnamefont {Brown}},
  \bibinfo {author} {\bibfnamefont {Y.}~\bibnamefont {Kim}}, \bibinfo {author}
  {\bibfnamefont {W.~G.}\ \bibnamefont {Lynch}}, \bibinfo {author}
  {\bibfnamefont {R.}~\bibnamefont {Michaels}}, \bibinfo {author}
  {\bibfnamefont {A.}~\bibnamefont {Ono}}, \bibinfo {author} {\bibfnamefont
  {J.}~\bibnamefont {Piekarewicz}}, \bibinfo {author} {\bibfnamefont {M.~B.}\
  \bibnamefont {Tsang}}, \ and\ \bibinfo {author} {\bibfnamefont {H.~H.}\
  \bibnamefont {Wolter}},\ }\href {\doibase 10.1088/0954-3899/41/9/093001}
  {\bibfield  {journal} {\bibinfo  {journal} {J. Phys. G}\ }\textbf {\bibinfo
  {volume} {41}},\ \bibinfo {pages} {093001} (\bibinfo {year}
  {2014})}\BibitemShut {NoStop}%
\bibitem [{\citenamefont {Thiel}\ \emph {et~al.}(2019)\citenamefont {Thiel},
  \citenamefont {Sfienti}, \citenamefont {Piekarewicz}, \citenamefont
  {Horowitz},\ and\ \citenamefont {Vanderhaeghen}}]{Thi19}%
  \BibitemOpen
  \bibfield  {author} {\bibinfo {author} {\bibfnamefont {M.}~\bibnamefont
  {Thiel}}, \bibinfo {author} {\bibfnamefont {C.}~\bibnamefont {Sfienti}},
  \bibinfo {author} {\bibfnamefont {J.}~\bibnamefont {Piekarewicz}}, \bibinfo
  {author} {\bibfnamefont {C.~J.}\ \bibnamefont {Horowitz}}, \ and\ \bibinfo
  {author} {\bibfnamefont {M.}~\bibnamefont {Vanderhaeghen}},\ }\href {\doibase
  10.1088/1361-6471/ab2c6d} {\bibfield  {journal} {\bibinfo  {journal} {J.
  Phys. G}\ }\textbf {\bibinfo {volume} {46}},\ \bibinfo {pages} {093003}
  (\bibinfo {year} {2019})}\BibitemShut {NoStop}%
\bibitem [{\citenamefont {Lattimer}\ and\ \citenamefont
  {Prakash}(2007)}]{Lattimer:2006xb}%
  \BibitemOpen
  \bibfield  {author} {\bibinfo {author} {\bibfnamefont {J.~M.}\ \bibnamefont
  {Lattimer}}\ and\ \bibinfo {author} {\bibfnamefont {M.}~\bibnamefont
  {Prakash}},\ }\href {\doibase https://doi.org/10.1016/j.physrep.2007.02.003}
  {\bibfield  {journal} {\bibinfo  {journal} {Phys. Rep.}\ }\textbf {\bibinfo
  {volume} {442}},\ \bibinfo {pages} {109} (\bibinfo {year}
  {2007})}\BibitemShut {NoStop}%
\bibitem [{\citenamefont {Horowitz}\ and\ \citenamefont
  {Piekarewicz}(2001{\natexlab{a}})}]{Horowitz:2000xj}%
  \BibitemOpen
  \bibfield  {author} {\bibinfo {author} {\bibfnamefont {C.~J.}\ \bibnamefont
  {Horowitz}}\ and\ \bibinfo {author} {\bibfnamefont {J.}~\bibnamefont
  {Piekarewicz}},\ }\href {\doibase 10.1103/PhysRevLett.86.5647} {\bibfield
  {journal} {\bibinfo  {journal} {Phys. Rev. Lett.}\ }\textbf {\bibinfo
  {volume} {86}},\ \bibinfo {pages} {5647} (\bibinfo {year}
  {2001}{\natexlab{a}})}\BibitemShut {NoStop}%
\bibitem [{\citenamefont {Horowitz}\ and\ \citenamefont
  {Piekarewicz}(2001{\natexlab{b}})}]{Horowitz:2001ya}%
  \BibitemOpen
  \bibfield  {author} {\bibinfo {author} {\bibfnamefont {C.~J.}\ \bibnamefont
  {Horowitz}}\ and\ \bibinfo {author} {\bibfnamefont {J.}~\bibnamefont
  {Piekarewicz}},\ }\href {\doibase 10.1103/PhysRevC.64.062802} {\bibfield
  {journal} {\bibinfo  {journal} {Phys. Rev. C}\ }\textbf {\bibinfo {volume}
  {64}},\ \bibinfo {pages} {062802(R)} (\bibinfo {year}
  {2001}{\natexlab{b}})}\BibitemShut {NoStop}%
\bibitem [{\citenamefont {Abrahamyan}\ \emph {et~al.}(2012)\citenamefont
  {Abrahamyan}, \citenamefont {Ahmed}, \citenamefont {Albataineh},
  \citenamefont {Aniol}, \citenamefont {Armstrong}, \citenamefont {Armstrong},
  \citenamefont {Averett}, \citenamefont {Babineau}, \citenamefont {Barbieri},
  \citenamefont {Bellini}, \citenamefont {Beminiwattha}, \citenamefont
  {Benesch}, \citenamefont {Benmokhtar}, \citenamefont {Bielarski},
  \citenamefont {Boeglin}, \citenamefont {Camsonne}, \citenamefont {Canan},
  \citenamefont {Carter}, \citenamefont {Cates}, \citenamefont {Chen},
  \citenamefont {Chen}, \citenamefont {Hen}, \citenamefont {Cusanno},
  \citenamefont {Dalton}, \citenamefont {De~Leo}, \citenamefont {de~Jager},
  \citenamefont {Deconinck}, \citenamefont {Decowski}, \citenamefont {Deng},
  \citenamefont {Deur}, \citenamefont {Dutta}, \citenamefont {Etile},
  \citenamefont {Flay}, \citenamefont {Franklin}, \citenamefont {Friend},
  \citenamefont {Frullani}, \citenamefont {Fuchey}, \citenamefont {Garibaldi},
  \citenamefont {Gasser}, \citenamefont {Gilman}, \citenamefont {Giusa},
  \citenamefont {Glamazdin}, \citenamefont {Gomez}, \citenamefont {Grames},
  \citenamefont {Gu}, \citenamefont {Hansen}, \citenamefont {Hansknecht},
  \citenamefont {Higinbotham}, \citenamefont {Holmes}, \citenamefont
  {Holmstrom}, \citenamefont {Horowitz}, \citenamefont {Hoskins}, \citenamefont
  {Huang}, \citenamefont {Hyde}, \citenamefont {Itard}, \citenamefont {Jen},
  \citenamefont {Jensen}, \citenamefont {Jin}, \citenamefont {Johnston},
  \citenamefont {Kelleher}, \citenamefont {Kliakhandler}, \citenamefont {King},
  \citenamefont {Kowalski}, \citenamefont {Kumar}, \citenamefont {Leacock},
  \citenamefont {Leckey}, \citenamefont {Lee}, \citenamefont {LeRose},
  \citenamefont {Lindgren}, \citenamefont {Liyanage}, \citenamefont {Lubinsky},
  \citenamefont {Mammei}, \citenamefont {Mammoliti}, \citenamefont
  {Margaziotis}, \citenamefont {Markowitz}, \citenamefont {McCreary},
  \citenamefont {McNulty}, \citenamefont {Mercado}, \citenamefont {Meziani},
  \citenamefont {Michaels}, \citenamefont {Mihovilovic}, \citenamefont
  {Muangma}, \citenamefont {Mu\~noz Camacho}, \citenamefont {Nanda},
  \citenamefont {Nelyubin}, \citenamefont {Nuruzzaman}, \citenamefont {Oh},
  \citenamefont {Palmer}, \citenamefont {Parno}, \citenamefont {Paschke},
  \citenamefont {Phillips}, \citenamefont {Poelker}, \citenamefont
  {Pomatsalyuk}, \citenamefont {Posik}, \citenamefont {Puckett}, \citenamefont
  {Quinn}, \citenamefont {Rakhman}, \citenamefont {Reimer}, \citenamefont
  {Riordan}, \citenamefont {Rogan}, \citenamefont {Ron}, \citenamefont {Russo},
  \citenamefont {Saenboonruang}, \citenamefont {Saha}, \citenamefont
  {Sawatzky}, \citenamefont {Shahinyan}, \citenamefont {Silwal}, \citenamefont
  {Sirca}, \citenamefont {Slifer}, \citenamefont {Solvignon}, \citenamefont
  {Souder}, \citenamefont {Sperduto}, \citenamefont {Subedi}, \citenamefont
  {Suleiman}, \citenamefont {Sulkosky}, \citenamefont {Sutera}, \citenamefont
  {Tobias}, \citenamefont {Troth}, \citenamefont {Urciuoli}, \citenamefont
  {Waidyawansa}, \citenamefont {Wang}, \citenamefont {Wexler}, \citenamefont
  {Wilson}, \citenamefont {Wojtsekhowski}, \citenamefont {Yan}, \citenamefont
  {Yao}, \citenamefont {Ye}, \citenamefont {Ye}, \citenamefont {Yim},
  \citenamefont {Zana}, \citenamefont {Zhan}, \citenamefont {Zhang},
  \citenamefont {Zhang}, \citenamefont {Zheng},\ and\ \citenamefont
  {Zhu}}]{PREXI}%
  \BibitemOpen
  \bibfield  {author} {\bibinfo {author} {\bibfnamefont {S.}~\bibnamefont
  {Abrahamyan}}, \bibinfo {author} {\bibfnamefont {Z.}~\bibnamefont {Ahmed}},
  \bibinfo {author} {\bibfnamefont {H.}~\bibnamefont {Albataineh}}, \bibinfo
  {author} {\bibfnamefont {K.}~\bibnamefont {Aniol}}, \bibinfo {author}
  {\bibfnamefont {D.~S.}\ \bibnamefont {Armstrong}}, \bibinfo {author}
  {\bibfnamefont {W.}~\bibnamefont {Armstrong}}, \bibinfo {author}
  {\bibfnamefont {T.}~\bibnamefont {Averett}}, \bibinfo {author} {\bibfnamefont
  {B.}~\bibnamefont {Babineau}}, \bibinfo {author} {\bibfnamefont
  {A.}~\bibnamefont {Barbieri}}, \bibinfo {author} {\bibfnamefont
  {V.}~\bibnamefont {Bellini}}, \bibinfo {author} {\bibfnamefont
  {R.}~\bibnamefont {Beminiwattha}}, \bibinfo {author} {\bibfnamefont
  {J.}~\bibnamefont {Benesch}}, \bibinfo {author} {\bibfnamefont
  {F.}~\bibnamefont {Benmokhtar}}, \bibinfo {author} {\bibfnamefont
  {T.}~\bibnamefont {Bielarski}}, \bibinfo {author} {\bibfnamefont
  {W.}~\bibnamefont {Boeglin}}, \bibinfo {author} {\bibfnamefont
  {A.}~\bibnamefont {Camsonne}}, \bibinfo {author} {\bibfnamefont
  {M.}~\bibnamefont {Canan}}, \bibinfo {author} {\bibfnamefont
  {P.}~\bibnamefont {Carter}}, \bibinfo {author} {\bibfnamefont {G.~D.}\
  \bibnamefont {Cates}}, \bibinfo {author} {\bibfnamefont {C.}~\bibnamefont
  {Chen}}, \bibinfo {author} {\bibfnamefont {J.-P.}\ \bibnamefont {Chen}},
  \bibinfo {author} {\bibfnamefont {O.}~\bibnamefont {Hen}}, \bibinfo {author}
  {\bibfnamefont {F.}~\bibnamefont {Cusanno}}, \bibinfo {author} {\bibfnamefont
  {M.~M.}\ \bibnamefont {Dalton}}, \bibinfo {author} {\bibfnamefont
  {R.}~\bibnamefont {De~Leo}}, \bibinfo {author} {\bibfnamefont
  {K.}~\bibnamefont {de~Jager}}, \bibinfo {author} {\bibfnamefont
  {W.}~\bibnamefont {Deconinck}}, \bibinfo {author} {\bibfnamefont
  {P.}~\bibnamefont {Decowski}}, \bibinfo {author} {\bibfnamefont
  {X.}~\bibnamefont {Deng}}, \bibinfo {author} {\bibfnamefont {A.}~\bibnamefont
  {Deur}}, \bibinfo {author} {\bibfnamefont {D.}~\bibnamefont {Dutta}},
  \bibinfo {author} {\bibfnamefont {A.}~\bibnamefont {Etile}}, \bibinfo
  {author} {\bibfnamefont {D.}~\bibnamefont {Flay}}, \bibinfo {author}
  {\bibfnamefont {G.~B.}\ \bibnamefont {Franklin}}, \bibinfo {author}
  {\bibfnamefont {M.}~\bibnamefont {Friend}}, \bibinfo {author} {\bibfnamefont
  {S.}~\bibnamefont {Frullani}}, \bibinfo {author} {\bibfnamefont
  {E.}~\bibnamefont {Fuchey}}, \bibinfo {author} {\bibfnamefont
  {F.}~\bibnamefont {Garibaldi}}, \bibinfo {author} {\bibfnamefont
  {E.}~\bibnamefont {Gasser}}, \bibinfo {author} {\bibfnamefont
  {R.}~\bibnamefont {Gilman}}, \bibinfo {author} {\bibfnamefont
  {A.}~\bibnamefont {Giusa}}, \bibinfo {author} {\bibfnamefont
  {A.}~\bibnamefont {Glamazdin}}, \bibinfo {author} {\bibfnamefont
  {J.}~\bibnamefont {Gomez}}, \bibinfo {author} {\bibfnamefont
  {J.}~\bibnamefont {Grames}}, \bibinfo {author} {\bibfnamefont
  {C.}~\bibnamefont {Gu}}, \bibinfo {author} {\bibfnamefont {O.}~\bibnamefont
  {Hansen}}, \bibinfo {author} {\bibfnamefont {J.}~\bibnamefont {Hansknecht}},
  \bibinfo {author} {\bibfnamefont {D.~W.}\ \bibnamefont {Higinbotham}},
  \bibinfo {author} {\bibfnamefont {R.~S.}\ \bibnamefont {Holmes}}, \bibinfo
  {author} {\bibfnamefont {T.}~\bibnamefont {Holmstrom}}, \bibinfo {author}
  {\bibfnamefont {C.~J.}\ \bibnamefont {Horowitz}}, \bibinfo {author}
  {\bibfnamefont {J.}~\bibnamefont {Hoskins}}, \bibinfo {author} {\bibfnamefont
  {J.}~\bibnamefont {Huang}}, \bibinfo {author} {\bibfnamefont {C.~E.}\
  \bibnamefont {Hyde}}, \bibinfo {author} {\bibfnamefont {F.}~\bibnamefont
  {Itard}}, \bibinfo {author} {\bibfnamefont {C.-M.}\ \bibnamefont {Jen}},
  \bibinfo {author} {\bibfnamefont {E.}~\bibnamefont {Jensen}}, \bibinfo
  {author} {\bibfnamefont {G.}~\bibnamefont {Jin}}, \bibinfo {author}
  {\bibfnamefont {S.}~\bibnamefont {Johnston}}, \bibinfo {author}
  {\bibfnamefont {A.}~\bibnamefont {Kelleher}}, \bibinfo {author}
  {\bibfnamefont {K.}~\bibnamefont {Kliakhandler}}, \bibinfo {author}
  {\bibfnamefont {P.~M.}\ \bibnamefont {King}}, \bibinfo {author}
  {\bibfnamefont {S.}~\bibnamefont {Kowalski}}, \bibinfo {author}
  {\bibfnamefont {K.~S.}\ \bibnamefont {Kumar}}, \bibinfo {author}
  {\bibfnamefont {J.}~\bibnamefont {Leacock}}, \bibinfo {author} {\bibfnamefont
  {J.}~\bibnamefont {Leckey}}, \bibinfo {author} {\bibfnamefont {J.~H.}\
  \bibnamefont {Lee}}, \bibinfo {author} {\bibfnamefont {J.~J.}\ \bibnamefont
  {LeRose}}, \bibinfo {author} {\bibfnamefont {R.}~\bibnamefont {Lindgren}},
  \bibinfo {author} {\bibfnamefont {N.}~\bibnamefont {Liyanage}}, \bibinfo
  {author} {\bibfnamefont {N.}~\bibnamefont {Lubinsky}}, \bibinfo {author}
  {\bibfnamefont {J.}~\bibnamefont {Mammei}}, \bibinfo {author} {\bibfnamefont
  {F.}~\bibnamefont {Mammoliti}}, \bibinfo {author} {\bibfnamefont {D.~J.}\
  \bibnamefont {Margaziotis}}, \bibinfo {author} {\bibfnamefont
  {P.}~\bibnamefont {Markowitz}}, \bibinfo {author} {\bibfnamefont
  {A.}~\bibnamefont {McCreary}}, \bibinfo {author} {\bibfnamefont
  {D.}~\bibnamefont {McNulty}}, \bibinfo {author} {\bibfnamefont
  {L.}~\bibnamefont {Mercado}}, \bibinfo {author} {\bibfnamefont {Z.-E.}\
  \bibnamefont {Meziani}}, \bibinfo {author} {\bibfnamefont {R.~W.}\
  \bibnamefont {Michaels}}, \bibinfo {author} {\bibfnamefont {M.}~\bibnamefont
  {Mihovilovic}}, \bibinfo {author} {\bibfnamefont {N.}~\bibnamefont
  {Muangma}}, \bibinfo {author} {\bibfnamefont {C.}~\bibnamefont {Mu\~noz
  Camacho}}, \bibinfo {author} {\bibfnamefont {S.}~\bibnamefont {Nanda}},
  \bibinfo {author} {\bibfnamefont {V.}~\bibnamefont {Nelyubin}}, \bibinfo
  {author} {\bibfnamefont {N.}~\bibnamefont {Nuruzzaman}}, \bibinfo {author}
  {\bibfnamefont {Y.}~\bibnamefont {Oh}}, \bibinfo {author} {\bibfnamefont
  {A.}~\bibnamefont {Palmer}}, \bibinfo {author} {\bibfnamefont
  {D.}~\bibnamefont {Parno}}, \bibinfo {author} {\bibfnamefont {K.~D.}\
  \bibnamefont {Paschke}}, \bibinfo {author} {\bibfnamefont {S.~K.}\
  \bibnamefont {Phillips}}, \bibinfo {author} {\bibfnamefont {B.}~\bibnamefont
  {Poelker}}, \bibinfo {author} {\bibfnamefont {R.}~\bibnamefont
  {Pomatsalyuk}}, \bibinfo {author} {\bibfnamefont {M.}~\bibnamefont {Posik}},
  \bibinfo {author} {\bibfnamefont {A.~J.~R.}\ \bibnamefont {Puckett}},
  \bibinfo {author} {\bibfnamefont {B.}~\bibnamefont {Quinn}}, \bibinfo
  {author} {\bibfnamefont {A.}~\bibnamefont {Rakhman}}, \bibinfo {author}
  {\bibfnamefont {P.~E.}\ \bibnamefont {Reimer}}, \bibinfo {author}
  {\bibfnamefont {S.}~\bibnamefont {Riordan}}, \bibinfo {author} {\bibfnamefont
  {P.}~\bibnamefont {Rogan}}, \bibinfo {author} {\bibfnamefont
  {G.}~\bibnamefont {Ron}}, \bibinfo {author} {\bibfnamefont {G.}~\bibnamefont
  {Russo}}, \bibinfo {author} {\bibfnamefont {K.}~\bibnamefont
  {Saenboonruang}}, \bibinfo {author} {\bibfnamefont {A.}~\bibnamefont {Saha}},
  \bibinfo {author} {\bibfnamefont {B.}~\bibnamefont {Sawatzky}}, \bibinfo
  {author} {\bibfnamefont {A.}~\bibnamefont {Shahinyan}}, \bibinfo {author}
  {\bibfnamefont {R.}~\bibnamefont {Silwal}}, \bibinfo {author} {\bibfnamefont
  {S.}~\bibnamefont {Sirca}}, \bibinfo {author} {\bibfnamefont
  {K.}~\bibnamefont {Slifer}}, \bibinfo {author} {\bibfnamefont
  {P.}~\bibnamefont {Solvignon}}, \bibinfo {author} {\bibfnamefont {P.~A.}\
  \bibnamefont {Souder}}, \bibinfo {author} {\bibfnamefont {M.~L.}\
  \bibnamefont {Sperduto}}, \bibinfo {author} {\bibfnamefont {R.}~\bibnamefont
  {Subedi}}, \bibinfo {author} {\bibfnamefont {R.}~\bibnamefont {Suleiman}},
  \bibinfo {author} {\bibfnamefont {V.}~\bibnamefont {Sulkosky}}, \bibinfo
  {author} {\bibfnamefont {C.~M.}\ \bibnamefont {Sutera}}, \bibinfo {author}
  {\bibfnamefont {W.~A.}\ \bibnamefont {Tobias}}, \bibinfo {author}
  {\bibfnamefont {W.}~\bibnamefont {Troth}}, \bibinfo {author} {\bibfnamefont
  {G.~M.}\ \bibnamefont {Urciuoli}}, \bibinfo {author} {\bibfnamefont
  {B.}~\bibnamefont {Waidyawansa}}, \bibinfo {author} {\bibfnamefont
  {D.}~\bibnamefont {Wang}}, \bibinfo {author} {\bibfnamefont {J.}~\bibnamefont
  {Wexler}}, \bibinfo {author} {\bibfnamefont {R.}~\bibnamefont {Wilson}},
  \bibinfo {author} {\bibfnamefont {B.}~\bibnamefont {Wojtsekhowski}}, \bibinfo
  {author} {\bibfnamefont {X.}~\bibnamefont {Yan}}, \bibinfo {author}
  {\bibfnamefont {H.}~\bibnamefont {Yao}}, \bibinfo {author} {\bibfnamefont
  {Y.}~\bibnamefont {Ye}}, \bibinfo {author} {\bibfnamefont {Z.}~\bibnamefont
  {Ye}}, \bibinfo {author} {\bibfnamefont {V.}~\bibnamefont {Yim}}, \bibinfo
  {author} {\bibfnamefont {L.}~\bibnamefont {Zana}}, \bibinfo {author}
  {\bibfnamefont {X.}~\bibnamefont {Zhan}}, \bibinfo {author} {\bibfnamefont
  {J.}~\bibnamefont {Zhang}}, \bibinfo {author} {\bibfnamefont
  {Y.}~\bibnamefont {Zhang}}, \bibinfo {author} {\bibfnamefont
  {X.}~\bibnamefont {Zheng}}, \ and\ \bibinfo {author} {\bibfnamefont
  {P.}~\bibnamefont {Zhu}} (\bibinfo {collaboration} {PREX Collaboration}),\
  }\href {\doibase 10.1103/PhysRevLett.108.112502} {\bibfield  {journal}
  {\bibinfo  {journal} {Phys. Rev. Lett.}\ }\textbf {\bibinfo {volume} {108}},\
  \bibinfo {pages} {112502} (\bibinfo {year} {2012})}\BibitemShut {NoStop}%
\bibitem [{\citenamefont {Adhikari}\ \emph {et~al.}(2021)\citenamefont
  {Adhikari}, \citenamefont {Albataineh}, \citenamefont {Androic},
  \citenamefont {Aniol}, \citenamefont {Armstrong}, \citenamefont {Averett},
  \citenamefont {Ayerbe~Gayoso}, \citenamefont {Barcus}, \citenamefont
  {Bellini}, \citenamefont {Beminiwattha}, \citenamefont {Benesch},
  \citenamefont {Bhatt}, \citenamefont {Bhatta~Pathak}, \citenamefont
  {Bhetuwal}, \citenamefont {Blaikie}, \citenamefont {Campagna}, \citenamefont
  {Camsonne}, \citenamefont {Cates}, \citenamefont {Chen}, \citenamefont
  {Clarke}, \citenamefont {Cornejo}, \citenamefont {Covrig~Dusa}, \citenamefont
  {Datta}, \citenamefont {Deshpande}, \citenamefont {Dutta}, \citenamefont
  {Feldman}, \citenamefont {Fuchey}, \citenamefont {Gal}, \citenamefont
  {Gaskell}, \citenamefont {Gautam}, \citenamefont {Gericke}, \citenamefont
  {Ghosh}, \citenamefont {Halilovic}, \citenamefont {Hansen}, \citenamefont
  {Hauenstein}, \citenamefont {Henry}, \citenamefont {Horowitz}, \citenamefont
  {Jantzi}, \citenamefont {Jian}, \citenamefont {Johnston}, \citenamefont
  {Jones}, \citenamefont {Karki}, \citenamefont {Katugampola}, \citenamefont
  {Keppel}, \citenamefont {King}, \citenamefont {King}, \citenamefont {Knauss},
  \citenamefont {Kumar}, \citenamefont {Kutz}, \citenamefont
  {Lashley-Colthirst}, \citenamefont {Leverick}, \citenamefont {Liu},
  \citenamefont {Liyange}, \citenamefont {Malace}, \citenamefont {Mammei},
  \citenamefont {Mammei}, \citenamefont {McCaughan}, \citenamefont {McNulty},
  \citenamefont {Meekins}, \citenamefont {Metts}, \citenamefont {Michaels},
  \citenamefont {Mondal}, \citenamefont {Napolitano}, \citenamefont {Narayan},
  \citenamefont {Nikolaev}, \citenamefont {Rashad}, \citenamefont {Owen},
  \citenamefont {Palatchi}, \citenamefont {Pan}, \citenamefont {Pandey},
  \citenamefont {Park}, \citenamefont {Paschke}, \citenamefont {Petrusky},
  \citenamefont {Pitt}, \citenamefont {Premathilake}, \citenamefont {Puckett},
  \citenamefont {Quinn}, \citenamefont {Radloff}, \citenamefont {Rahman},
  \citenamefont {Rathnayake}, \citenamefont {Reed}, \citenamefont {Reimer},
  \citenamefont {Richards}, \citenamefont {Riordan}, \citenamefont {Roblin},
  \citenamefont {Seeds}, \citenamefont {Shahinyan}, \citenamefont {Souder},
  \citenamefont {Tang}, \citenamefont {Thiel}, \citenamefont {Tian},
  \citenamefont {Urciuoli}, \citenamefont {Wertz}, \citenamefont
  {Wojtsekhowski}, \citenamefont {Yale}, \citenamefont {Ye}, \citenamefont
  {Yoon}, \citenamefont {Zec}, \citenamefont {Zhang}, \citenamefont {Zhang},\
  and\ \citenamefont {Zheng}}]{PREXII}%
  \BibitemOpen
  \bibfield  {author} {\bibinfo {author} {\bibfnamefont {D.}~\bibnamefont
  {Adhikari}}, \bibinfo {author} {\bibfnamefont {H.}~\bibnamefont
  {Albataineh}}, \bibinfo {author} {\bibfnamefont {D.}~\bibnamefont {Androic}},
  \bibinfo {author} {\bibfnamefont {K.}~\bibnamefont {Aniol}}, \bibinfo
  {author} {\bibfnamefont {D.~S.}\ \bibnamefont {Armstrong}}, \bibinfo {author}
  {\bibfnamefont {T.}~\bibnamefont {Averett}}, \bibinfo {author} {\bibfnamefont
  {C.}~\bibnamefont {Ayerbe~Gayoso}}, \bibinfo {author} {\bibfnamefont
  {S.}~\bibnamefont {Barcus}}, \bibinfo {author} {\bibfnamefont
  {V.}~\bibnamefont {Bellini}}, \bibinfo {author} {\bibfnamefont {R.~S.}\
  \bibnamefont {Beminiwattha}}, \bibinfo {author} {\bibfnamefont {J.~F.}\
  \bibnamefont {Benesch}}, \bibinfo {author} {\bibfnamefont {H.}~\bibnamefont
  {Bhatt}}, \bibinfo {author} {\bibfnamefont {D.}~\bibnamefont
  {Bhatta~Pathak}}, \bibinfo {author} {\bibfnamefont {D.}~\bibnamefont
  {Bhetuwal}}, \bibinfo {author} {\bibfnamefont {B.}~\bibnamefont {Blaikie}},
  \bibinfo {author} {\bibfnamefont {Q.}~\bibnamefont {Campagna}}, \bibinfo
  {author} {\bibfnamefont {A.}~\bibnamefont {Camsonne}}, \bibinfo {author}
  {\bibfnamefont {G.~D.}\ \bibnamefont {Cates}}, \bibinfo {author}
  {\bibfnamefont {Y.}~\bibnamefont {Chen}}, \bibinfo {author} {\bibfnamefont
  {C.}~\bibnamefont {Clarke}}, \bibinfo {author} {\bibfnamefont {J.~C.}\
  \bibnamefont {Cornejo}}, \bibinfo {author} {\bibfnamefont {S.}~\bibnamefont
  {Covrig~Dusa}}, \bibinfo {author} {\bibfnamefont {P.}~\bibnamefont {Datta}},
  \bibinfo {author} {\bibfnamefont {A.}~\bibnamefont {Deshpande}}, \bibinfo
  {author} {\bibfnamefont {D.}~\bibnamefont {Dutta}}, \bibinfo {author}
  {\bibfnamefont {C.}~\bibnamefont {Feldman}}, \bibinfo {author} {\bibfnamefont
  {E.}~\bibnamefont {Fuchey}}, \bibinfo {author} {\bibfnamefont
  {C.}~\bibnamefont {Gal}}, \bibinfo {author} {\bibfnamefont {D.}~\bibnamefont
  {Gaskell}}, \bibinfo {author} {\bibfnamefont {T.}~\bibnamefont {Gautam}},
  \bibinfo {author} {\bibfnamefont {M.}~\bibnamefont {Gericke}}, \bibinfo
  {author} {\bibfnamefont {C.}~\bibnamefont {Ghosh}}, \bibinfo {author}
  {\bibfnamefont {I.}~\bibnamefont {Halilovic}}, \bibinfo {author}
  {\bibfnamefont {J.-O.}\ \bibnamefont {Hansen}}, \bibinfo {author}
  {\bibfnamefont {F.}~\bibnamefont {Hauenstein}}, \bibinfo {author}
  {\bibfnamefont {W.}~\bibnamefont {Henry}}, \bibinfo {author} {\bibfnamefont
  {C.~J.}\ \bibnamefont {Horowitz}}, \bibinfo {author} {\bibfnamefont
  {C.}~\bibnamefont {Jantzi}}, \bibinfo {author} {\bibfnamefont
  {S.}~\bibnamefont {Jian}}, \bibinfo {author} {\bibfnamefont {S.}~\bibnamefont
  {Johnston}}, \bibinfo {author} {\bibfnamefont {D.~C.}\ \bibnamefont {Jones}},
  \bibinfo {author} {\bibfnamefont {B.}~\bibnamefont {Karki}}, \bibinfo
  {author} {\bibfnamefont {S.}~\bibnamefont {Katugampola}}, \bibinfo {author}
  {\bibfnamefont {C.}~\bibnamefont {Keppel}}, \bibinfo {author} {\bibfnamefont
  {P.~M.}\ \bibnamefont {King}}, \bibinfo {author} {\bibfnamefont {D.~E.}\
  \bibnamefont {King}}, \bibinfo {author} {\bibfnamefont {M.}~\bibnamefont
  {Knauss}}, \bibinfo {author} {\bibfnamefont {K.~S.}\ \bibnamefont {Kumar}},
  \bibinfo {author} {\bibfnamefont {T.}~\bibnamefont {Kutz}}, \bibinfo {author}
  {\bibfnamefont {N.}~\bibnamefont {Lashley-Colthirst}}, \bibinfo {author}
  {\bibfnamefont {G.}~\bibnamefont {Leverick}}, \bibinfo {author}
  {\bibfnamefont {H.}~\bibnamefont {Liu}}, \bibinfo {author} {\bibfnamefont
  {N.}~\bibnamefont {Liyange}}, \bibinfo {author} {\bibfnamefont
  {S.}~\bibnamefont {Malace}}, \bibinfo {author} {\bibfnamefont
  {R.}~\bibnamefont {Mammei}}, \bibinfo {author} {\bibfnamefont
  {J.}~\bibnamefont {Mammei}}, \bibinfo {author} {\bibfnamefont
  {M.}~\bibnamefont {McCaughan}}, \bibinfo {author} {\bibfnamefont
  {D.}~\bibnamefont {McNulty}}, \bibinfo {author} {\bibfnamefont
  {D.}~\bibnamefont {Meekins}}, \bibinfo {author} {\bibfnamefont
  {C.}~\bibnamefont {Metts}}, \bibinfo {author} {\bibfnamefont
  {R.}~\bibnamefont {Michaels}}, \bibinfo {author} {\bibfnamefont {M.~M.}\
  \bibnamefont {Mondal}}, \bibinfo {author} {\bibfnamefont {J.}~\bibnamefont
  {Napolitano}}, \bibinfo {author} {\bibfnamefont {A.}~\bibnamefont {Narayan}},
  \bibinfo {author} {\bibfnamefont {D.}~\bibnamefont {Nikolaev}}, \bibinfo
  {author} {\bibfnamefont {M.~N.~H.}\ \bibnamefont {Rashad}}, \bibinfo {author}
  {\bibfnamefont {V.}~\bibnamefont {Owen}}, \bibinfo {author} {\bibfnamefont
  {C.}~\bibnamefont {Palatchi}}, \bibinfo {author} {\bibfnamefont
  {J.}~\bibnamefont {Pan}}, \bibinfo {author} {\bibfnamefont {B.}~\bibnamefont
  {Pandey}}, \bibinfo {author} {\bibfnamefont {S.}~\bibnamefont {Park}},
  \bibinfo {author} {\bibfnamefont {K.~D.}\ \bibnamefont {Paschke}}, \bibinfo
  {author} {\bibfnamefont {M.}~\bibnamefont {Petrusky}}, \bibinfo {author}
  {\bibfnamefont {M.~L.}\ \bibnamefont {Pitt}}, \bibinfo {author}
  {\bibfnamefont {S.}~\bibnamefont {Premathilake}}, \bibinfo {author}
  {\bibfnamefont {A.~J.~R.}\ \bibnamefont {Puckett}}, \bibinfo {author}
  {\bibfnamefont {B.}~\bibnamefont {Quinn}}, \bibinfo {author} {\bibfnamefont
  {R.}~\bibnamefont {Radloff}}, \bibinfo {author} {\bibfnamefont
  {S.}~\bibnamefont {Rahman}}, \bibinfo {author} {\bibfnamefont
  {A.}~\bibnamefont {Rathnayake}}, \bibinfo {author} {\bibfnamefont {B.~T.}\
  \bibnamefont {Reed}}, \bibinfo {author} {\bibfnamefont {P.~E.}\ \bibnamefont
  {Reimer}}, \bibinfo {author} {\bibfnamefont {R.}~\bibnamefont {Richards}},
  \bibinfo {author} {\bibfnamefont {S.}~\bibnamefont {Riordan}}, \bibinfo
  {author} {\bibfnamefont {Y.}~\bibnamefont {Roblin}}, \bibinfo {author}
  {\bibfnamefont {S.}~\bibnamefont {Seeds}}, \bibinfo {author} {\bibfnamefont
  {A.}~\bibnamefont {Shahinyan}}, \bibinfo {author} {\bibfnamefont
  {P.}~\bibnamefont {Souder}}, \bibinfo {author} {\bibfnamefont
  {L.}~\bibnamefont {Tang}}, \bibinfo {author} {\bibfnamefont {M.}~\bibnamefont
  {Thiel}}, \bibinfo {author} {\bibfnamefont {Y.}~\bibnamefont {Tian}},
  \bibinfo {author} {\bibfnamefont {G.~M.}\ \bibnamefont {Urciuoli}}, \bibinfo
  {author} {\bibfnamefont {E.~W.}\ \bibnamefont {Wertz}}, \bibinfo {author}
  {\bibfnamefont {B.}~\bibnamefont {Wojtsekhowski}}, \bibinfo {author}
  {\bibfnamefont {B.}~\bibnamefont {Yale}}, \bibinfo {author} {\bibfnamefont
  {T.}~\bibnamefont {Ye}}, \bibinfo {author} {\bibfnamefont {A.}~\bibnamefont
  {Yoon}}, \bibinfo {author} {\bibfnamefont {A.}~\bibnamefont {Zec}}, \bibinfo
  {author} {\bibfnamefont {W.}~\bibnamefont {Zhang}}, \bibinfo {author}
  {\bibfnamefont {J.}~\bibnamefont {Zhang}}, \ and\ \bibinfo {author}
  {\bibfnamefont {X.}~\bibnamefont {Zheng}} (\bibinfo {collaboration} {PREX
  Collaboration}),\ }\href {\doibase 10.1103/PhysRevLett.126.172502} {\bibfield
   {journal} {\bibinfo  {journal} {Phys. Rev. Lett.}\ }\textbf {\bibinfo
  {volume} {126}},\ \bibinfo {pages} {172502} (\bibinfo {year}
  {2021})}\BibitemShut {NoStop}%
\bibitem [{\citenamefont {Zenihiro}\ \emph {et~al.}(2010)\citenamefont
  {Zenihiro}, \citenamefont {Sakaguchi}, \citenamefont {Murakami},
  \citenamefont {Yosoi}, \citenamefont {Yasuda}, \citenamefont {Terashima},
  \citenamefont {Iwao}, \citenamefont {Takeda}, \citenamefont {Itoh},
  \citenamefont {Yoshida},\ and\ \citenamefont {Uchida}}]{Zen10}%
  \BibitemOpen
  \bibfield  {author} {\bibinfo {author} {\bibfnamefont {J.}~\bibnamefont
  {Zenihiro}}, \bibinfo {author} {\bibfnamefont {H.}~\bibnamefont {Sakaguchi}},
  \bibinfo {author} {\bibfnamefont {T.}~\bibnamefont {Murakami}}, \bibinfo
  {author} {\bibfnamefont {M.}~\bibnamefont {Yosoi}}, \bibinfo {author}
  {\bibfnamefont {Y.}~\bibnamefont {Yasuda}}, \bibinfo {author} {\bibfnamefont
  {S.}~\bibnamefont {Terashima}}, \bibinfo {author} {\bibfnamefont
  {Y.}~\bibnamefont {Iwao}}, \bibinfo {author} {\bibfnamefont {H.}~\bibnamefont
  {Takeda}}, \bibinfo {author} {\bibfnamefont {M.}~\bibnamefont {Itoh}},
  \bibinfo {author} {\bibfnamefont {H.~P.}\ \bibnamefont {Yoshida}}, \ and\
  \bibinfo {author} {\bibfnamefont {M.}~\bibnamefont {Uchida}},\ }\href
  {\doibase 10.1103/PhysRevC.82.044611} {\bibfield  {journal} {\bibinfo
  {journal} {Phys. Rev. C}\ }\textbf {\bibinfo {volume} {82}},\ \bibinfo
  {pages} {044611} (\bibinfo {year} {2010})}\BibitemShut {NoStop}%
\bibitem [{\citenamefont {Trzci\ifmmode~\acute{n}\else \'{n}\fi{}ska}\ \emph
  {et~al.}(2001)\citenamefont {Trzci\ifmmode~\acute{n}\else \'{n}\fi{}ska},
  \citenamefont {Jastrz\ifmmode~\mbox{\c{e}}\else \c{e}\fi{}bski},
  \citenamefont {Lubi\ifmmode~\acute{n}\else \'{n}\fi{}ski}, \citenamefont
  {Hartmann}, \citenamefont {Schmidt}, \citenamefont {von Egidy},\ and\
  \citenamefont {K\l{}os}}]{Trz01}%
  \BibitemOpen
  \bibfield  {author} {\bibinfo {author} {\bibfnamefont {A.}~\bibnamefont
  {Trzci\ifmmode~\acute{n}\else \'{n}\fi{}ska}}, \bibinfo {author}
  {\bibfnamefont {J.}~\bibnamefont {Jastrz\ifmmode~\mbox{\c{e}}\else
  \c{e}\fi{}bski}}, \bibinfo {author} {\bibfnamefont {P.}~\bibnamefont
  {Lubi\ifmmode~\acute{n}\else \'{n}\fi{}ski}}, \bibinfo {author}
  {\bibfnamefont {F.~J.}\ \bibnamefont {Hartmann}}, \bibinfo {author}
  {\bibfnamefont {R.}~\bibnamefont {Schmidt}}, \bibinfo {author} {\bibfnamefont
  {T.}~\bibnamefont {von Egidy}}, \ and\ \bibinfo {author} {\bibfnamefont
  {B.}~\bibnamefont {K\l{}os}},\ }\href {\doibase
  10.1103/PhysRevLett.87.082501} {\bibfield  {journal} {\bibinfo  {journal}
  {Phys. Rev. Lett.}\ }\textbf {\bibinfo {volume} {87}},\ \bibinfo {pages}
  {082501} (\bibinfo {year} {2001})}\BibitemShut {NoStop}%
\bibitem [{\citenamefont {K\l{}os}\ \emph {et~al.}(2007)\citenamefont
  {K\l{}os}, \citenamefont {Trzci\ifmmode~\acute{n}\else \'{n}\fi{}ska},
  \citenamefont {Jastrz\c{e}bski}, \citenamefont {Czosnyka}, \citenamefont
  {Kisieli\ifmmode~\acute{n}\else \'{n}\fi{}ski}, \citenamefont
  {Lubi\ifmmode~\acute{n}\else \'{n}\fi{}ski}, \citenamefont {Napiorkowski},
  \citenamefont {Pie\ifmmode~\acute{n}\else \'{n}\fi{}kowski}, \citenamefont
  {Hartmann}, \citenamefont {Ketzer}, \citenamefont {Ring}, \citenamefont
  {Schmidt}, \citenamefont {von Egidy}, \citenamefont
  {Smola\ifmmode~\acute{n}\else \'{n}\fi{}czuk}, \citenamefont {Wycech},
  \citenamefont {Gulda}, \citenamefont {Kurcewicz}, \citenamefont {Widmann},\
  and\ \citenamefont {Brown}}]{Klo07}%
  \BibitemOpen
  \bibfield  {author} {\bibinfo {author} {\bibfnamefont {B.}~\bibnamefont
  {K\l{}os}}, \bibinfo {author} {\bibfnamefont {A.}~\bibnamefont
  {Trzci\ifmmode~\acute{n}\else \'{n}\fi{}ska}}, \bibinfo {author}
  {\bibfnamefont {J.}~\bibnamefont {Jastrz\c{e}bski}}, \bibinfo {author}
  {\bibfnamefont {T.}~\bibnamefont {Czosnyka}}, \bibinfo {author}
  {\bibfnamefont {M.}~\bibnamefont {Kisieli\ifmmode~\acute{n}\else
  \'{n}\fi{}ski}}, \bibinfo {author} {\bibfnamefont {P.}~\bibnamefont
  {Lubi\ifmmode~\acute{n}\else \'{n}\fi{}ski}}, \bibinfo {author}
  {\bibfnamefont {P.}~\bibnamefont {Napiorkowski}}, \bibinfo {author}
  {\bibfnamefont {L.}~\bibnamefont {Pie\ifmmode~\acute{n}\else
  \'{n}\fi{}kowski}}, \bibinfo {author} {\bibfnamefont {F.~J.}\ \bibnamefont
  {Hartmann}}, \bibinfo {author} {\bibfnamefont {B.}~\bibnamefont {Ketzer}},
  \bibinfo {author} {\bibfnamefont {P.}~\bibnamefont {Ring}}, \bibinfo {author}
  {\bibfnamefont {R.}~\bibnamefont {Schmidt}}, \bibinfo {author} {\bibfnamefont
  {T.}~\bibnamefont {von Egidy}}, \bibinfo {author} {\bibfnamefont
  {R.}~\bibnamefont {Smola\ifmmode~\acute{n}\else \'{n}\fi{}czuk}}, \bibinfo
  {author} {\bibfnamefont {S.}~\bibnamefont {Wycech}}, \bibinfo {author}
  {\bibfnamefont {K.}~\bibnamefont {Gulda}}, \bibinfo {author} {\bibfnamefont
  {W.}~\bibnamefont {Kurcewicz}}, \bibinfo {author} {\bibfnamefont
  {E.}~\bibnamefont {Widmann}}, \ and\ \bibinfo {author} {\bibfnamefont
  {B.~A.}\ \bibnamefont {Brown}},\ }\href {\doibase 10.1103/PhysRevC.76.014311}
  {\bibfield  {journal} {\bibinfo  {journal} {Phys. Rev. C}\ }\textbf {\bibinfo
  {volume} {76}},\ \bibinfo {pages} {014311} (\bibinfo {year}
  {2007})}\BibitemShut {NoStop}%
\bibitem [{\citenamefont {Tamii}\ \emph {et~al.}(2011)\citenamefont {Tamii},
  \citenamefont {Poltoratska}, \citenamefont {von Neumann-Cosel}, \citenamefont
  {Fujita}, \citenamefont {Adachi}, \citenamefont {Bertulani}, \citenamefont
  {Carter}, \citenamefont {Dozono}, \citenamefont {Fujita}, \citenamefont
  {Fujita}, \citenamefont {Hatanaka}, \citenamefont {Ishikawa}, \citenamefont
  {Itoh}, \citenamefont {Kawabata}, \citenamefont {Kalmykov}, \citenamefont
  {Krumbholz}, \citenamefont {Litvinova}, \citenamefont {Matsubara},
  \citenamefont {Nakanishi}, \citenamefont {Neveling}, \citenamefont {Okamura},
  \citenamefont {Ong}, \citenamefont {\"Ozel-Tashenov}, \citenamefont
  {Ponomarev}, \citenamefont {Richter}, \citenamefont {Rubio}, \citenamefont
  {Sakaguchi}, \citenamefont {Sakemi}, \citenamefont {Sasamoto}, \citenamefont
  {Shimbara}, \citenamefont {Shimizu}, \citenamefont {Smit}, \citenamefont
  {Suzuki}, \citenamefont {Tameshige}, \citenamefont {Wambach}, \citenamefont
  {Yamada}, \citenamefont {Yosoi},\ and\ \citenamefont {Zenihiro}}]{Tam11}%
  \BibitemOpen
  \bibfield  {author} {\bibinfo {author} {\bibfnamefont {A.}~\bibnamefont
  {Tamii}}, \bibinfo {author} {\bibfnamefont {I.}~\bibnamefont {Poltoratska}},
  \bibinfo {author} {\bibfnamefont {P.}~\bibnamefont {von Neumann-Cosel}},
  \bibinfo {author} {\bibfnamefont {Y.}~\bibnamefont {Fujita}}, \bibinfo
  {author} {\bibfnamefont {T.}~\bibnamefont {Adachi}}, \bibinfo {author}
  {\bibfnamefont {C.~A.}\ \bibnamefont {Bertulani}}, \bibinfo {author}
  {\bibfnamefont {J.}~\bibnamefont {Carter}}, \bibinfo {author} {\bibfnamefont
  {M.}~\bibnamefont {Dozono}}, \bibinfo {author} {\bibfnamefont
  {H.}~\bibnamefont {Fujita}}, \bibinfo {author} {\bibfnamefont
  {K.}~\bibnamefont {Fujita}}, \bibinfo {author} {\bibfnamefont
  {K.}~\bibnamefont {Hatanaka}}, \bibinfo {author} {\bibfnamefont
  {D.}~\bibnamefont {Ishikawa}}, \bibinfo {author} {\bibfnamefont
  {M.}~\bibnamefont {Itoh}}, \bibinfo {author} {\bibfnamefont {T.}~\bibnamefont
  {Kawabata}}, \bibinfo {author} {\bibfnamefont {Y.}~\bibnamefont {Kalmykov}},
  \bibinfo {author} {\bibfnamefont {A.~M.}\ \bibnamefont {Krumbholz}}, \bibinfo
  {author} {\bibfnamefont {E.}~\bibnamefont {Litvinova}}, \bibinfo {author}
  {\bibfnamefont {H.}~\bibnamefont {Matsubara}}, \bibinfo {author}
  {\bibfnamefont {K.}~\bibnamefont {Nakanishi}}, \bibinfo {author}
  {\bibfnamefont {R.}~\bibnamefont {Neveling}}, \bibinfo {author}
  {\bibfnamefont {H.}~\bibnamefont {Okamura}}, \bibinfo {author} {\bibfnamefont
  {H.~J.}\ \bibnamefont {Ong}}, \bibinfo {author} {\bibfnamefont
  {B.}~\bibnamefont {\"Ozel-Tashenov}}, \bibinfo {author} {\bibfnamefont
  {V.~Y.}\ \bibnamefont {Ponomarev}}, \bibinfo {author} {\bibfnamefont
  {A.}~\bibnamefont {Richter}}, \bibinfo {author} {\bibfnamefont
  {B.}~\bibnamefont {Rubio}}, \bibinfo {author} {\bibfnamefont
  {H.}~\bibnamefont {Sakaguchi}}, \bibinfo {author} {\bibfnamefont
  {Y.}~\bibnamefont {Sakemi}}, \bibinfo {author} {\bibfnamefont
  {Y.}~\bibnamefont {Sasamoto}}, \bibinfo {author} {\bibfnamefont
  {Y.}~\bibnamefont {Shimbara}}, \bibinfo {author} {\bibfnamefont
  {Y.}~\bibnamefont {Shimizu}}, \bibinfo {author} {\bibfnamefont {F.~D.}\
  \bibnamefont {Smit}}, \bibinfo {author} {\bibfnamefont {T.}~\bibnamefont
  {Suzuki}}, \bibinfo {author} {\bibfnamefont {Y.}~\bibnamefont {Tameshige}},
  \bibinfo {author} {\bibfnamefont {J.}~\bibnamefont {Wambach}}, \bibinfo
  {author} {\bibfnamefont {R.}~\bibnamefont {Yamada}}, \bibinfo {author}
  {\bibfnamefont {M.}~\bibnamefont {Yosoi}}, \ and\ \bibinfo {author}
  {\bibfnamefont {J.}~\bibnamefont {Zenihiro}},\ }\href {\doibase
  10.1103/PhysRevLett.107.062502} {\bibfield  {journal} {\bibinfo  {journal}
  {Phys. Rev. Lett.}\ }\textbf {\bibinfo {volume} {107}},\ \bibinfo {pages}
  {062502} (\bibinfo {year} {2011})}\BibitemShut {NoStop}%
\bibitem [{\citenamefont {Reinhard}\ and\ \citenamefont
  {Nazarewicz}(2010)}]{Reinhard:2010wz}%
  \BibitemOpen
  \bibfield  {author} {\bibinfo {author} {\bibfnamefont {P.-G.}\ \bibnamefont
  {Reinhard}}\ and\ \bibinfo {author} {\bibfnamefont {W.}~\bibnamefont
  {Nazarewicz}},\ }\href {\doibase 10.1103/PhysRevC.81.051303} {\bibfield
  {journal} {\bibinfo  {journal} {Phys. Rev. C}\ }\textbf {\bibinfo {volume}
  {81}},\ \bibinfo {pages} {051303(R)} (\bibinfo {year} {2010})}\BibitemShut
  {NoStop}%
\bibitem [{\citenamefont {Schrack}\ \emph {et~al.}(1962)\citenamefont
  {Schrack}, \citenamefont {Leiss},\ and\ \citenamefont
  {Penner}}]{PhysRev.127.1772}%
  \BibitemOpen
  \bibfield  {author} {\bibinfo {author} {\bibfnamefont {R.~A.}\ \bibnamefont
  {Schrack}}, \bibinfo {author} {\bibfnamefont {J.~E.}\ \bibnamefont {Leiss}},
  \ and\ \bibinfo {author} {\bibfnamefont {S.}~\bibnamefont {Penner}},\ }\href
  {\doibase 10.1103/PhysRev.127.1772} {\bibfield  {journal} {\bibinfo
  {journal} {Phys. Rev.}\ }\textbf {\bibinfo {volume} {127}},\ \bibinfo {pages}
  {1772} (\bibinfo {year} {1962})}\BibitemShut {NoStop}%
\bibitem [{\citenamefont {Leiss}\ and\ \citenamefont
  {Schrack}(1958)}]{RevModPhys.30.456}%
  \BibitemOpen
  \bibfield  {author} {\bibinfo {author} {\bibfnamefont {J.~E.}\ \bibnamefont
  {Leiss}}\ and\ \bibinfo {author} {\bibfnamefont {R.~A.}\ \bibnamefont
  {Schrack}},\ }\href {\doibase 10.1103/RevModPhys.30.456} {\bibfield
  {journal} {\bibinfo  {journal} {Rev. Mod. Phys.}\ }\textbf {\bibinfo {volume}
  {30}},\ \bibinfo {pages} {456} (\bibinfo {year} {1958})}\BibitemShut
  {NoStop}%
\bibitem [{\citenamefont {Drechsel}\ \emph {et~al.}(1999)\citenamefont
  {Drechsel}, \citenamefont {Tiator}, \citenamefont {Kamalov},\ and\
  \citenamefont {Yang}}]{NPA660}%
  \BibitemOpen
  \bibfield  {author} {\bibinfo {author} {\bibfnamefont {D.}~\bibnamefont
  {Drechsel}}, \bibinfo {author} {\bibfnamefont {L.}~\bibnamefont {Tiator}},
  \bibinfo {author} {\bibfnamefont {S.}~\bibnamefont {Kamalov}}, \ and\
  \bibinfo {author} {\bibfnamefont {S.~N.}\ \bibnamefont {Yang}},\ }\href
  {\doibase https://doi.org/10.1016/S0375-9474(99)00412-1} {\bibfield
  {journal} {\bibinfo  {journal} {Nucl. Phys.}\ }\textbf {\bibinfo {volume}
  {A660}},\ \bibinfo {pages} {423 } (\bibinfo {year} {1999})}\BibitemShut
  {NoStop}%
\bibitem [{\citenamefont {Fricke}\ \emph {et~al.}(1995)\citenamefont {Fricke},
  \citenamefont {Bernhardt}, \citenamefont {Heilig}, \citenamefont {Schaller},
  \citenamefont {Schellenberg}, \citenamefont {Shera},\ and\ \citenamefont
  {Dejager}}]{Fri95}%
  \BibitemOpen
  \bibfield  {author} {\bibinfo {author} {\bibfnamefont {G.}~\bibnamefont
  {Fricke}}, \bibinfo {author} {\bibfnamefont {C.}~\bibnamefont {Bernhardt}},
  \bibinfo {author} {\bibfnamefont {K.}~\bibnamefont {Heilig}}, \bibinfo
  {author} {\bibfnamefont {L.}~\bibnamefont {Schaller}}, \bibinfo {author}
  {\bibfnamefont {L.}~\bibnamefont {Schellenberg}}, \bibinfo {author}
  {\bibfnamefont {E.}~\bibnamefont {Shera}}, \ and\ \bibinfo {author}
  {\bibfnamefont {C.}~\bibnamefont {Dejager}},\ }\href {\doibase
  https://doi.org/10.1006/adnd.1995.1007} {\bibfield  {journal} {\bibinfo
  {journal} {At. Data Nucl. Data Tables}\ }\textbf {\bibinfo {volume} {60}},\
  \bibinfo {pages} {177} (\bibinfo {year} {1995})}\BibitemShut {NoStop}%
\bibitem [{\citenamefont {Krusche}\ \emph {et~al.}(2002)\citenamefont
  {Krusche}, \citenamefont {Ahrens}, \citenamefont {Beck}, \citenamefont
  {Kamalov}, \citenamefont {Metag}, \citenamefont {Owens},\ and\ \citenamefont
  {Ströher}}]{KRUSCHE2002287}%
  \BibitemOpen
  \bibfield  {author} {\bibinfo {author} {\bibfnamefont {B.}~\bibnamefont
  {Krusche}}, \bibinfo {author} {\bibfnamefont {J.}~\bibnamefont {Ahrens}},
  \bibinfo {author} {\bibfnamefont {R.}~\bibnamefont {Beck}}, \bibinfo {author}
  {\bibfnamefont {S.}~\bibnamefont {Kamalov}}, \bibinfo {author} {\bibfnamefont
  {V.}~\bibnamefont {Metag}}, \bibinfo {author} {\bibfnamefont
  {R.}~\bibnamefont {Owens}}, \ and\ \bibinfo {author} {\bibfnamefont
  {H.}~\bibnamefont {Ströher}},\ }\href {\doibase
  https://doi.org/10.1016/S0370-2693(01)01503-9} {\bibfield  {journal}
  {\bibinfo  {journal} {Phys. Lett.}\ }\textbf {\bibinfo {volume} {B526}},\
  \bibinfo {pages} {287 } (\bibinfo {year} {2002})}\BibitemShut {NoStop}%
\bibitem [{\citenamefont {Tarbert}\ \emph {et~al.}(2014)\citenamefont
  {Tarbert}, \citenamefont {Watts}, \citenamefont {Glazier}, \citenamefont
  {Aguar}, \citenamefont {Ahrens}, \citenamefont {Annand}, \citenamefont
  {Arends}, \citenamefont {Beck}, \citenamefont {Bekrenev}, \citenamefont
  {Boillat}, \citenamefont {Braghieri}, \citenamefont {Branford}, \citenamefont
  {Briscoe}, \citenamefont {Brudvik}, \citenamefont {Cherepnya}, \citenamefont
  {Codling}, \citenamefont {Downie}, \citenamefont {Foehl}, \citenamefont
  {Grabmayr}, \citenamefont {Gregor}, \citenamefont {Heid}, \citenamefont
  {Hornidge}, \citenamefont {Jahn}, \citenamefont {Kashevarov}, \citenamefont
  {Knezevic}, \citenamefont {Kondratiev}, \citenamefont {Korolija},
  \citenamefont {Kotulla}, \citenamefont {Krambrich}, \citenamefont {Krusche},
  \citenamefont {Lang}, \citenamefont {Lisin}, \citenamefont {Livingston},
  \citenamefont {Lugert}, \citenamefont {MacGregor}, \citenamefont {Manley},
  \citenamefont {Martinez}, \citenamefont {McGeorge}, \citenamefont
  {Mekterovic}, \citenamefont {Metag}, \citenamefont {Nefkens}, \citenamefont
  {Nikolaev}, \citenamefont {Novotny}, \citenamefont {Owens}, \citenamefont
  {Pedroni}, \citenamefont {Polonski}, \citenamefont {Prakhov}, \citenamefont
  {Price}, \citenamefont {Rosner}, \citenamefont {Rost}, \citenamefont
  {Rostomyan}, \citenamefont {Schadmand}, \citenamefont {Schumann},
  \citenamefont {Sober}, \citenamefont {Starostin}, \citenamefont {Supek},
  \citenamefont {Thomas}, \citenamefont {Unverzagt}, \citenamefont {Walcher},
  \citenamefont {Zana},\ and\ \citenamefont {Zehr}}]{PhysRevLett112_242502}%
  \BibitemOpen
  \bibfield  {author} {\bibinfo {author} {\bibfnamefont {C.~M.}\ \bibnamefont
  {Tarbert}}, \bibinfo {author} {\bibfnamefont {D.~P.}\ \bibnamefont {Watts}},
  \bibinfo {author} {\bibfnamefont {D.~I.}\ \bibnamefont {Glazier}}, \bibinfo
  {author} {\bibfnamefont {P.}~\bibnamefont {Aguar}}, \bibinfo {author}
  {\bibfnamefont {J.}~\bibnamefont {Ahrens}}, \bibinfo {author} {\bibfnamefont
  {J.~R.~M.}\ \bibnamefont {Annand}}, \bibinfo {author} {\bibfnamefont {H.~J.}\
  \bibnamefont {Arends}}, \bibinfo {author} {\bibfnamefont {R.}~\bibnamefont
  {Beck}}, \bibinfo {author} {\bibfnamefont {V.}~\bibnamefont {Bekrenev}},
  \bibinfo {author} {\bibfnamefont {B.}~\bibnamefont {Boillat}}, \bibinfo
  {author} {\bibfnamefont {A.}~\bibnamefont {Braghieri}}, \bibinfo {author}
  {\bibfnamefont {D.}~\bibnamefont {Branford}}, \bibinfo {author}
  {\bibfnamefont {W.~J.}\ \bibnamefont {Briscoe}}, \bibinfo {author}
  {\bibfnamefont {J.}~\bibnamefont {Brudvik}}, \bibinfo {author} {\bibfnamefont
  {S.}~\bibnamefont {Cherepnya}}, \bibinfo {author} {\bibfnamefont
  {R.}~\bibnamefont {Codling}}, \bibinfo {author} {\bibfnamefont {E.~J.}\
  \bibnamefont {Downie}}, \bibinfo {author} {\bibfnamefont {K.}~\bibnamefont
  {Foehl}}, \bibinfo {author} {\bibfnamefont {P.}~\bibnamefont {Grabmayr}},
  \bibinfo {author} {\bibfnamefont {R.}~\bibnamefont {Gregor}}, \bibinfo
  {author} {\bibfnamefont {E.}~\bibnamefont {Heid}}, \bibinfo {author}
  {\bibfnamefont {D.}~\bibnamefont {Hornidge}}, \bibinfo {author}
  {\bibfnamefont {O.}~\bibnamefont {Jahn}}, \bibinfo {author} {\bibfnamefont
  {V.~L.}\ \bibnamefont {Kashevarov}}, \bibinfo {author} {\bibfnamefont
  {A.}~\bibnamefont {Knezevic}}, \bibinfo {author} {\bibfnamefont
  {R.}~\bibnamefont {Kondratiev}}, \bibinfo {author} {\bibfnamefont
  {M.}~\bibnamefont {Korolija}}, \bibinfo {author} {\bibfnamefont
  {M.}~\bibnamefont {Kotulla}}, \bibinfo {author} {\bibfnamefont
  {D.}~\bibnamefont {Krambrich}}, \bibinfo {author} {\bibfnamefont
  {B.}~\bibnamefont {Krusche}}, \bibinfo {author} {\bibfnamefont
  {M.}~\bibnamefont {Lang}}, \bibinfo {author} {\bibfnamefont {V.}~\bibnamefont
  {Lisin}}, \bibinfo {author} {\bibfnamefont {K.}~\bibnamefont {Livingston}},
  \bibinfo {author} {\bibfnamefont {S.}~\bibnamefont {Lugert}}, \bibinfo
  {author} {\bibfnamefont {I.~J.~D.}\ \bibnamefont {MacGregor}}, \bibinfo
  {author} {\bibfnamefont {D.~M.}\ \bibnamefont {Manley}}, \bibinfo {author}
  {\bibfnamefont {M.}~\bibnamefont {Martinez}}, \bibinfo {author}
  {\bibfnamefont {J.~C.}\ \bibnamefont {McGeorge}}, \bibinfo {author}
  {\bibfnamefont {D.}~\bibnamefont {Mekterovic}}, \bibinfo {author}
  {\bibfnamefont {V.}~\bibnamefont {Metag}}, \bibinfo {author} {\bibfnamefont
  {B.~M.~K.}\ \bibnamefont {Nefkens}}, \bibinfo {author} {\bibfnamefont
  {A.}~\bibnamefont {Nikolaev}}, \bibinfo {author} {\bibfnamefont
  {R.}~\bibnamefont {Novotny}}, \bibinfo {author} {\bibfnamefont {R.~O.}\
  \bibnamefont {Owens}}, \bibinfo {author} {\bibfnamefont {P.}~\bibnamefont
  {Pedroni}}, \bibinfo {author} {\bibfnamefont {A.}~\bibnamefont {Polonski}},
  \bibinfo {author} {\bibfnamefont {S.~N.}\ \bibnamefont {Prakhov}}, \bibinfo
  {author} {\bibfnamefont {J.~W.}\ \bibnamefont {Price}}, \bibinfo {author}
  {\bibfnamefont {G.}~\bibnamefont {Rosner}}, \bibinfo {author} {\bibfnamefont
  {M.}~\bibnamefont {Rost}}, \bibinfo {author} {\bibfnamefont {T.}~\bibnamefont
  {Rostomyan}}, \bibinfo {author} {\bibfnamefont {S.}~\bibnamefont
  {Schadmand}}, \bibinfo {author} {\bibfnamefont {S.}~\bibnamefont {Schumann}},
  \bibinfo {author} {\bibfnamefont {D.}~\bibnamefont {Sober}}, \bibinfo
  {author} {\bibfnamefont {A.}~\bibnamefont {Starostin}}, \bibinfo {author}
  {\bibfnamefont {I.}~\bibnamefont {Supek}}, \bibinfo {author} {\bibfnamefont
  {A.}~\bibnamefont {Thomas}}, \bibinfo {author} {\bibfnamefont
  {M.}~\bibnamefont {Unverzagt}}, \bibinfo {author} {\bibfnamefont
  {T.}~\bibnamefont {Walcher}}, \bibinfo {author} {\bibfnamefont
  {L.}~\bibnamefont {Zana}}, \ and\ \bibinfo {author} {\bibfnamefont
  {F.}~\bibnamefont {Zehr}} (\bibinfo {collaboration} {Crystal Ball at MAMI and
  A2 Collaboration}),\ }\href {\doibase 10.1103/PhysRevLett.112.242502}
  {\bibfield  {journal} {\bibinfo  {journal} {Phys. Rev. Lett.}\ }\textbf
  {\bibinfo {volume} {112}},\ \bibinfo {pages} {242502} (\bibinfo {year}
  {2014})}\BibitemShut {NoStop}%
\bibitem [{\citenamefont {Miller}(2019)}]{Mil19}%
  \BibitemOpen
  \bibfield  {author} {\bibinfo {author} {\bibfnamefont {G.~A.}\ \bibnamefont
  {Miller}},\ }\href {\doibase 10.1103/PhysRevC.100.044608} {\bibfield
  {journal} {\bibinfo  {journal} {Phys. Rev. C}\ }\textbf {\bibinfo {volume}
  {100}},\ \bibinfo {pages} {044608} (\bibinfo {year} {2019})}\BibitemShut
  {NoStop}%
\bibitem [{\citenamefont {Peters}\ \emph {et~al.}(1998)\citenamefont {Peters},
  \citenamefont {Lenske},\ and\ \citenamefont {Mosel}}]{PLM98}%
  \BibitemOpen
  \bibfield  {author} {\bibinfo {author} {\bibfnamefont {W.}~\bibnamefont
  {Peters}}, \bibinfo {author} {\bibfnamefont {H.}~\bibnamefont {Lenske}}, \
  and\ \bibinfo {author} {\bibfnamefont {U.}~\bibnamefont {Mosel}},\ }\href
  {\doibase https://doi.org/10.1016/S0375-9474(98)00446-1} {\bibfield
  {journal} {\bibinfo  {journal} {Nucl. Phys. A}\ }\textbf {\bibinfo {volume}
  {640}},\ \bibinfo {pages} {89} (\bibinfo {year} {1998})}\BibitemShut
  {NoStop}%
\bibitem [{\citenamefont {Chew}\ \emph {et~al.}(1957)\citenamefont {Chew},
  \citenamefont {Goldberger}, \citenamefont {Low},\ and\ \citenamefont
  {Nambu}}]{CGLN57}%
  \BibitemOpen
  \bibfield  {author} {\bibinfo {author} {\bibfnamefont {G.~F.}\ \bibnamefont
  {Chew}}, \bibinfo {author} {\bibfnamefont {M.~L.}\ \bibnamefont
  {Goldberger}}, \bibinfo {author} {\bibfnamefont {F.~E.}\ \bibnamefont {Low}},
  \ and\ \bibinfo {author} {\bibfnamefont {Y.}~\bibnamefont {Nambu}},\ }\href
  {\doibase 10.1103/PhysRev.106.1345} {\bibfield  {journal} {\bibinfo
  {journal} {Phys. Rev.}\ }\textbf {\bibinfo {volume} {106}},\ \bibinfo {pages}
  {1345} (\bibinfo {year} {1957})}\BibitemShut {NoStop}%
\bibitem [{\citenamefont {Drechsel}\ \emph {et~al.}(2007)\citenamefont
  {Drechsel}, \citenamefont {Kamalov},\ and\ \citenamefont {Tiator}}]{MAID}%
  \BibitemOpen
  \bibfield  {author} {\bibinfo {author} {\bibfnamefont {D.}~\bibnamefont
  {Drechsel}}, \bibinfo {author} {\bibfnamefont {S.~S.}\ \bibnamefont
  {Kamalov}}, \ and\ \bibinfo {author} {\bibfnamefont {L.}~\bibnamefont
  {Tiator}},\ }\href {\doibase 10.1140/epja/i2007-10490-6} {\bibfield
  {journal} {\bibinfo  {journal} {Euro. Phys. J. A}\ }\textbf {\bibinfo
  {volume} {34}},\ \bibinfo {pages} {69} (\bibinfo {year} {2007})}\BibitemShut
  {NoStop}%
\bibitem [{\citenamefont {Gmitro}\ \emph {et~al.}(1985)\citenamefont {Gmitro},
  \citenamefont {Kvasil},\ and\ \citenamefont {Mach}}]{PhysRevC.31.1349}%
  \BibitemOpen
  \bibfield  {author} {\bibinfo {author} {\bibfnamefont {M.}~\bibnamefont
  {Gmitro}}, \bibinfo {author} {\bibfnamefont {J.}~\bibnamefont {Kvasil}}, \
  and\ \bibinfo {author} {\bibfnamefont {R.}~\bibnamefont {Mach}},\ }\href
  {\doibase 10.1103/PhysRevC.31.1349} {\bibfield  {journal} {\bibinfo
  {journal} {Phys. Rev. C}\ }\textbf {\bibinfo {volume} {31}},\ \bibinfo
  {pages} {1349} (\bibinfo {year} {1985})}\BibitemShut {NoStop}%
\bibitem [{\citenamefont {Carr}\ \emph {et~al.}(1982)\citenamefont {Carr},
  \citenamefont {McManus},\ and\ \citenamefont
  {Stricker-Bauer}}]{PhysRevC25_952}%
  \BibitemOpen
  \bibfield  {author} {\bibinfo {author} {\bibfnamefont {J.~A.}\ \bibnamefont
  {Carr}}, \bibinfo {author} {\bibfnamefont {H.}~\bibnamefont {McManus}}, \
  and\ \bibinfo {author} {\bibfnamefont {K.}~\bibnamefont {Stricker-Bauer}},\
  }\href {\doibase 10.1103/PhysRevC.25.952} {\bibfield  {journal} {\bibinfo
  {journal} {Phys. Rev. C}\ }\textbf {\bibinfo {volume} {25}},\ \bibinfo
  {pages} {952} (\bibinfo {year} {1982})}\BibitemShut {NoStop}%
\bibitem [{\citenamefont {Ericson}\ and\ \citenamefont
  {Ericson}(1966)}]{ERICSON1966323}%
  \BibitemOpen
  \bibfield  {author} {\bibinfo {author} {\bibfnamefont {M.}~\bibnamefont
  {Ericson}}\ and\ \bibinfo {author} {\bibfnamefont {T.}~\bibnamefont
  {Ericson}},\ }\href {\doibase https://doi.org/10.1016/0003-4916(66)90302-2}
  {\bibfield  {journal} {\bibinfo  {journal} {Ann. Phys.}\ }\textbf {\bibinfo
  {volume} {36}},\ \bibinfo {pages} {323 } (\bibinfo {year}
  {1966})}\BibitemShut {NoStop}%
\bibitem [{\citenamefont {Chamon}\ \emph {et~al.}(2002)\citenamefont {Chamon},
  \citenamefont {Carlson}, \citenamefont {Gasques}, \citenamefont {Pereira},
  \citenamefont {De~Conti}, \citenamefont {Alvarez}, \citenamefont {Hussein},
  \citenamefont {C\^andido~Ribeiro}, \citenamefont {Rossi},\ and\ \citenamefont
  {Silva}}]{PhysRevC.66.014610}%
  \BibitemOpen
  \bibfield  {author} {\bibinfo {author} {\bibfnamefont {L.~C.}\ \bibnamefont
  {Chamon}}, \bibinfo {author} {\bibfnamefont {B.~V.}\ \bibnamefont {Carlson}},
  \bibinfo {author} {\bibfnamefont {L.~R.}\ \bibnamefont {Gasques}}, \bibinfo
  {author} {\bibfnamefont {D.}~\bibnamefont {Pereira}}, \bibinfo {author}
  {\bibfnamefont {C.}~\bibnamefont {De~Conti}}, \bibinfo {author}
  {\bibfnamefont {M.~A.~G.}\ \bibnamefont {Alvarez}}, \bibinfo {author}
  {\bibfnamefont {M.~S.}\ \bibnamefont {Hussein}}, \bibinfo {author}
  {\bibfnamefont {M.~A.}\ \bibnamefont {C\^andido~Ribeiro}}, \bibinfo {author}
  {\bibfnamefont {E.~S.}\ \bibnamefont {Rossi}}, \ and\ \bibinfo {author}
  {\bibfnamefont {C.~P.}\ \bibnamefont {Silva}},\ }\href {\doibase
  10.1103/PhysRevC.66.014610} {\bibfield  {journal} {\bibinfo  {journal} {Phys.
  Rev. C}\ }\textbf {\bibinfo {volume} {66}},\ \bibinfo {pages} {014610}
  (\bibinfo {year} {2002})}\BibitemShut {NoStop}%
\bibitem [{\citenamefont {{De Jager}}\ \emph {et~al.}(1974)\citenamefont {{De
  Jager}}, \citenamefont {{De Vries}},\ and\ \citenamefont {{De
  Vries}}}]{DJDV74}%
  \BibitemOpen
  \bibfield  {author} {\bibinfo {author} {\bibfnamefont {C.}~\bibnamefont {{De
  Jager}}}, \bibinfo {author} {\bibfnamefont {H.}~\bibnamefont {{De Vries}}}, \
  and\ \bibinfo {author} {\bibfnamefont {C.}~\bibnamefont {{De Vries}}},\
  }\href {\doibase https://doi.org/10.1016/S0092-640X(74)80002-1} {\bibfield
  {journal} {\bibinfo  {journal} {At. Data Nucl. Data Tables}\ }\textbf
  {\bibinfo {volume} {14}},\ \bibinfo {pages} {479 } (\bibinfo {year}
  {1974})},\ \bibinfo {note} {nuclear Charge and Moment
  Distributions}\BibitemShut {NoStop}%
\bibitem [{\citenamefont {Dreher}\ \emph {et~al.}(1974)\citenamefont {Dreher},
  \citenamefont {Friedrich}, \citenamefont {Merle}, \citenamefont {Rothhaas},\
  and\ \citenamefont {Lührs}}]{DREHER1974219}%
  \BibitemOpen
  \bibfield  {author} {\bibinfo {author} {\bibfnamefont {B.}~\bibnamefont
  {Dreher}}, \bibinfo {author} {\bibfnamefont {J.}~\bibnamefont {Friedrich}},
  \bibinfo {author} {\bibfnamefont {K.}~\bibnamefont {Merle}}, \bibinfo
  {author} {\bibfnamefont {H.}~\bibnamefont {Rothhaas}}, \ and\ \bibinfo
  {author} {\bibfnamefont {G.}~\bibnamefont {Lührs}},\ }\href {\doibase
  https://doi.org/10.1016/0375-9474(74)90189-4} {\bibfield  {journal} {\bibinfo
   {journal} {Nucl. Phys.}\ }\textbf {\bibinfo {volume} {A235}},\ \bibinfo
  {pages} {219 } (\bibinfo {year} {1974})}\BibitemShut {NoStop}%
\bibitem [{\citenamefont {Todd-Rutel}\ and\ \citenamefont
  {Piekarewicz}(2005)}]{PhysRevLett.95.122501}%
  \BibitemOpen
  \bibfield  {author} {\bibinfo {author} {\bibfnamefont {B.~G.}\ \bibnamefont
  {Todd-Rutel}}\ and\ \bibinfo {author} {\bibfnamefont {J.}~\bibnamefont
  {Piekarewicz}},\ }\href {\doibase 10.1103/PhysRevLett.95.122501} {\bibfield
  {journal} {\bibinfo  {journal} {Phys. Rev. Lett.}\ }\textbf {\bibinfo
  {volume} {95}},\ \bibinfo {pages} {122501} (\bibinfo {year}
  {2005})}\BibitemShut {NoStop}%
\bibitem [{\citenamefont {Hagen}\ \emph {et~al.}(2016)\citenamefont {Hagen},
  \citenamefont {Ekstr\"om}, \citenamefont {Forss\'en}, \citenamefont {Jansen},
  \citenamefont {Nazarewicz}, \citenamefont {Papenbrock}, \citenamefont
  {Wendt}, \citenamefont {Bacca}, \citenamefont {Barnea}, \citenamefont
  {Carlsson}, \citenamefont {Drischler}, \citenamefont {Hebeler}, \citenamefont
  {Hjorth-Jensen}, \citenamefont {Miorelli}, \citenamefont {Orlandini},
  \citenamefont {Schwenk},\ and\ \citenamefont {Simonis}}]{Hag16}%
  \BibitemOpen
  \bibfield  {author} {\bibinfo {author} {\bibfnamefont {G.}~\bibnamefont
  {Hagen}}, \bibinfo {author} {\bibfnamefont {A.}~\bibnamefont {Ekstr\"om}},
  \bibinfo {author} {\bibfnamefont {C.}~\bibnamefont {Forss\'en}}, \bibinfo
  {author} {\bibfnamefont {G.~R.}\ \bibnamefont {Jansen}}, \bibinfo {author}
  {\bibfnamefont {W.}~\bibnamefont {Nazarewicz}}, \bibinfo {author}
  {\bibfnamefont {T.}~\bibnamefont {Papenbrock}}, \bibinfo {author}
  {\bibfnamefont {K.~A.}\ \bibnamefont {Wendt}}, \bibinfo {author}
  {\bibfnamefont {S.}~\bibnamefont {Bacca}}, \bibinfo {author} {\bibfnamefont
  {N.}~\bibnamefont {Barnea}}, \bibinfo {author} {\bibfnamefont
  {B.}~\bibnamefont {Carlsson}}, \bibinfo {author} {\bibfnamefont
  {C.}~\bibnamefont {Drischler}}, \bibinfo {author} {\bibfnamefont
  {K.}~\bibnamefont {Hebeler}}, \bibinfo {author} {\bibfnamefont
  {M.}~\bibnamefont {Hjorth-Jensen}}, \bibinfo {author} {\bibfnamefont
  {M.}~\bibnamefont {Miorelli}}, \bibinfo {author} {\bibfnamefont
  {G.}~\bibnamefont {Orlandini}}, \bibinfo {author} {\bibfnamefont
  {A.}~\bibnamefont {Schwenk}}, \ and\ \bibinfo {author} {\bibfnamefont
  {J.}~\bibnamefont {Simonis}},\ }\href {\doibase
  https://doi.org/10.1038/nphys3529} {\bibfield  {journal} {\bibinfo  {journal}
  {Nat. Phys.}\ }\textbf {\bibinfo {volume} {12}},\ \bibinfo {pages} {186}
  (\bibinfo {year} {2016})}\BibitemShut {NoStop}%
\bibitem [{\citenamefont {Hagen}(2016)}]{Hag16p}%
  \BibitemOpen
  \bibfield  {author} {\bibinfo {author} {\bibfnamefont {G.}~\bibnamefont
  {Hagen}},\ }\href@noop {} {}\bibinfo {howpublished} {(private communication)}
  (\bibinfo {year} {2016})\BibitemShut {NoStop}%
\bibitem [{\citenamefont {Chumbalov}\ \emph {et~al.}(1987)\citenamefont
  {Chumbalov}, \citenamefont {Eramzhyan},\ and\ \citenamefont
  {Kamalov}}]{ZPA.328.195}%
  \BibitemOpen
  \bibfield  {author} {\bibinfo {author} {\bibfnamefont {A.~A.}\ \bibnamefont
  {Chumbalov}}, \bibinfo {author} {\bibfnamefont {R.~A.}\ \bibnamefont
  {Eramzhyan}}, \ and\ \bibinfo {author} {\bibfnamefont {S.~S.}\ \bibnamefont
  {Kamalov}},\ }\href {\doibase 10.1007/BF01290662} {\bibfield  {journal}
  {\bibinfo  {journal} {Z. Phys. A}\ }\textbf {\bibinfo {volume} {328}},\
  \bibinfo {pages} {195} (\bibinfo {year} {1987})}\BibitemShut {NoStop}%
\bibitem [{\citenamefont {Krusche}(2018)}]{Kru18}%
  \BibitemOpen
  \bibfield  {author} {\bibinfo {author} {\bibfnamefont {B.}~\bibnamefont
  {Krusche}},\ }\href@noop {} {}\bibinfo {howpublished} {(private
  communication)} (\bibinfo {year} {2018})\BibitemShut {NoStop}%
\bibitem [{\citenamefont {Kaiser}\ \emph {et~al.}(2008)\citenamefont {Kaiser},
  \citenamefont {Aulenbacher}, \citenamefont {Chubarov}, \citenamefont {Dehn},
  \citenamefont {Euteneuer}, \citenamefont {Hagenbuck}, \citenamefont {Herr},
  \citenamefont {Jankowiak}, \citenamefont {Jennewein}, \citenamefont
  {Kreidel}, \citenamefont {Ludwig-Mertin}, \citenamefont {Negrazus},
  \citenamefont {Ratschow}, \citenamefont {Schumann}, \citenamefont {Seidl},
  \citenamefont {Stephan},\ and\ \citenamefont {Thomas}}]{MAMI}%
  \BibitemOpen
  \bibfield  {author} {\bibinfo {author} {\bibfnamefont {K.-H.}\ \bibnamefont
  {Kaiser}}, \bibinfo {author} {\bibfnamefont {K.}~\bibnamefont {Aulenbacher}},
  \bibinfo {author} {\bibfnamefont {O.}~\bibnamefont {Chubarov}}, \bibinfo
  {author} {\bibfnamefont {M.}~\bibnamefont {Dehn}}, \bibinfo {author}
  {\bibfnamefont {H.}~\bibnamefont {Euteneuer}}, \bibinfo {author}
  {\bibfnamefont {F.}~\bibnamefont {Hagenbuck}}, \bibinfo {author}
  {\bibfnamefont {R.}~\bibnamefont {Herr}}, \bibinfo {author} {\bibfnamefont
  {A.}~\bibnamefont {Jankowiak}}, \bibinfo {author} {\bibfnamefont
  {P.}~\bibnamefont {Jennewein}}, \bibinfo {author} {\bibfnamefont {H.-J.}\
  \bibnamefont {Kreidel}}, \bibinfo {author} {\bibfnamefont {U.}~\bibnamefont
  {Ludwig-Mertin}}, \bibinfo {author} {\bibfnamefont {M.}~\bibnamefont
  {Negrazus}}, \bibinfo {author} {\bibfnamefont {S.}~\bibnamefont {Ratschow}},
  \bibinfo {author} {\bibfnamefont {S.}~\bibnamefont {Schumann}}, \bibinfo
  {author} {\bibfnamefont {M.}~\bibnamefont {Seidl}}, \bibinfo {author}
  {\bibfnamefont {G.}~\bibnamefont {Stephan}}, \ and\ \bibinfo {author}
  {\bibfnamefont {A.}~\bibnamefont {Thomas}},\ }\href {\doibase
  https://doi.org/10.1016/j.nima.2008.05.018} {\bibfield  {journal} {\bibinfo
  {journal} {Nucl. Instrument. Methods A}\ }\textbf {\bibinfo {volume} {593}},\
  \bibinfo {pages} {159 } (\bibinfo {year} {2008})}\BibitemShut {NoStop}%
\bibitem [{\citenamefont {Todd-Rutel}\ \emph {et~al.}(2004)\citenamefont
  {Todd-Rutel}, \citenamefont {Piekarewicz},\ and\ \citenamefont
  {Cottle}}]{Todd-Rutel:2004llz}%
  \BibitemOpen
  \bibfield  {author} {\bibinfo {author} {\bibfnamefont {B.~G.}\ \bibnamefont
  {Todd-Rutel}}, \bibinfo {author} {\bibfnamefont {J.}~\bibnamefont
  {Piekarewicz}}, \ and\ \bibinfo {author} {\bibfnamefont {P.~D.}\ \bibnamefont
  {Cottle}},\ }\href {\doibase 10.1103/PhysRevC.69.021301} {\bibfield
  {journal} {\bibinfo  {journal} {Phys. Rev. C}\ }\textbf {\bibinfo {volume}
  {69}},\ \bibinfo {pages} {021301(R)} (\bibinfo {year} {2004})}\BibitemShut
  {NoStop}%
\bibitem [{\citenamefont {Grasso}\ \emph {et~al.}(2009)\citenamefont {Grasso},
  \citenamefont {Gaudefroy}, \citenamefont {Khan}, \citenamefont {Nik\ifmmode
  \check{s}\else \v{s}\fi{}i\ifmmode~\acute{c}\else \'{c}\fi{}}, \citenamefont
  {Piekarewicz}, \citenamefont {Sorlin}, \citenamefont {Van~Giai},\ and\
  \citenamefont {Vretenar}}]{Grasso:2009zza}%
  \BibitemOpen
  \bibfield  {author} {\bibinfo {author} {\bibfnamefont {M.}~\bibnamefont
  {Grasso}}, \bibinfo {author} {\bibfnamefont {L.}~\bibnamefont {Gaudefroy}},
  \bibinfo {author} {\bibfnamefont {E.}~\bibnamefont {Khan}}, \bibinfo {author}
  {\bibfnamefont {T.}~\bibnamefont {Nik\ifmmode \check{s}\else
  \v{s}\fi{}i\ifmmode~\acute{c}\else \'{c}\fi{}}}, \bibinfo {author}
  {\bibfnamefont {J.}~\bibnamefont {Piekarewicz}}, \bibinfo {author}
  {\bibfnamefont {O.}~\bibnamefont {Sorlin}}, \bibinfo {author} {\bibfnamefont
  {N.}~\bibnamefont {Van~Giai}}, \ and\ \bibinfo {author} {\bibfnamefont
  {D.}~\bibnamefont {Vretenar}},\ }\href {\doibase 10.1103/PhysRevC.79.034318}
  {\bibfield  {journal} {\bibinfo  {journal} {Phys. Rev. C}\ }\textbf {\bibinfo
  {volume} {79}},\ \bibinfo {pages} {034318} (\bibinfo {year}
  {2009})}\BibitemShut {NoStop}%
\bibitem [{\citenamefont {Piekarewicz}\ \emph {et~al.}(2016)\citenamefont
  {Piekarewicz}, \citenamefont {Linero}, \citenamefont {Giuliani},\ and\
  \citenamefont {Chicken}}]{Piekarewicz:2016vbn}%
  \BibitemOpen
  \bibfield  {author} {\bibinfo {author} {\bibfnamefont {J.}~\bibnamefont
  {Piekarewicz}}, \bibinfo {author} {\bibfnamefont {A.~R.}\ \bibnamefont
  {Linero}}, \bibinfo {author} {\bibfnamefont {P.}~\bibnamefont {Giuliani}}, \
  and\ \bibinfo {author} {\bibfnamefont {E.}~\bibnamefont {Chicken}},\ }\href
  {\doibase 10.1103/PhysRevC.94.034316} {\bibfield  {journal} {\bibinfo
  {journal} {Phys. Rev. C}\ }\textbf {\bibinfo {volume} {94}},\ \bibinfo
  {pages} {034316} (\bibinfo {year} {2016})}\BibitemShut {NoStop}%
\end{thebibliography}
%

\end{document}